\documentclass[aps,pre,amsmath,amssymb,amsfonts,preprint
]{revtex4-1}

\usepackage{graphicx}
\usepackage{dcolumn}
\usepackage{bm}

\begin{document}

\preprint{APS}

\title{Variational Schemes and Geometric Simulations for a Hydrodynamic-Electrodynamic Model of Surface Plasmon Polaritons}

\author{Qiang~Chen}
 \email{cq0405@ustc.edu.cn}
\affiliation{State Key Laboratory of Complex Electromagnetic Environment Effects on Electronics 
and Information System, Luoyang, Henan 471000, China}
\affiliation{University of Science and Technology of China, Hefei, Anhui 230026, China}
\author{Lifei~Geng}
\affiliation{State Key Laboratory of Complex Electromagnetic Environment Effects on Electronics 
and Information System, Luoyang, Henan 471000, China}
\author{Xiang~Chen}
\affiliation{State Key Laboratory of Complex Electromagnetic Environment Effects on Electronics 
and Information System, Luoyang, Henan 471000, China}
\author{Xiaojun~Hao}
\affiliation{State Key Laboratory of Complex Electromagnetic Environment Effects on Electronics 
and Information System, Luoyang, Henan 471000, China}
\author{Chuanchuan~Wang}
\affiliation{State Key Laboratory of Complex Electromagnetic Environment Effects on Electronics 
and Information System, Luoyang, Henan 471000, China}
\author{Xiaoyang~Wang}
\affiliation{State Key Laboratory of Complex Electromagnetic Environment Effects on Electronics 
and Information System, Luoyang, Henan 471000, China}

\date{\today}

\begin{abstract}
A class of variational schemes for the hydrodynamic-electrodynamic model of lossless free electron gas in a quasi-neutral 
background is developed for high-quality simulations of surface plasmon polaritons. The Lagrangian density of lossless 
free electron gas with a self-consistent electromagnetic field is established, and the dynamical equations with the 
associated constraints are obtained via a variational principle. Based on discrete exterior calculus, the action functional 
of this system is discretized and minimized to obtain the discrete dynamics. Newton-Raphson iteration and the biconjugate 
gradient stabilized method are equipped as a hybrid nonlinear-linear algebraic solver. Instead of discretizing the partial 
differential equations, the variational schemes have better numerical properties in secular simulations, as they preserve 
the discrete Lagrangian symplectic structure, gauge symmetry, and general energy-momentum density. Two numerical experiments 
were performed. The numerical results reproduce characteristic dispersion relations of bulk plasmons and surface plasmon 
polaritons, and the numerical errors of conserved quantities in all experiments are bounded by a small value after long 
term simulations.
\begin{description}
\item[PACS number(s)]
73.20.Mf, 11.10.Ef, 45.20.Jj, 02.40.Yy, 03.50.-z
\end{description}
\end{abstract}


\maketitle

\section{Introduction\label{sec:1}}

In the past two decades, there have been impressive developments and significant advancement in applications of Surface 
Plasmon Polaritons (SPPs), bringing many new ideas into traditional electromagnetics and optics, such as the lithography 
beyond the diffraction limit, chip-scale photonic circuits, plasmonic metasurfaces, bio-photonics, etc. \cite{Ritchie,
Barnes,Ozbay,Zia,Hendry,Kim,Atwater,Boltasseva,Valev,Fang}. In the field of metal optics, plasmonics focuses on the 
collective motions of free electron gas in metal with self-consistent and external electromagnetic fields whose 
first-principle model is the classical particle-field theory \cite{Ritchie,Barnes,Ozbay}. Direct applications of the 
first-principle model in macroscopic simulations face many obstacles, such as the nonlinearity, the multi-scale, and the 
huge degrees-of-freedom. As a simplification, linearized phenomenological models, e.g., the Drude-Lorentz (DL) model, 
are widely used to describe macroscopic plasmonic phenomena \cite{Barnes,Ozbay}. In a mesoscopic context, kinetic and 
hydrodynamic descriptions are basic physical models of plasmonics, which involve both the dynamics of free electron gas 
and an electromagnetic field. Therefore, high-quality numerical schemes and simulations based on the hydrodynamic model 
are necessary in plasmonic research.

The physics of SPPs can be described by the free electron gas model whose dynamical equations are hydrodynamic and 
Maxwell's equations \cite{Ritchie}. For Maxwell's equations, many numerical schemes, such as the Finite-Difference 
Time-Domain (FDTD) method, the Finite Element (FE) method, and the Method of Moments (MoM) have been developed, that are 
widely used in modern electromagnetic engineering, Radio Frequency (RF) and microwave engineering, terahertz 
engineering, optical engineering, metamaterial design, accelerator design, fusion engineering, radio astrophysics, 
geophysics, and even biomedicine \cite{Yee,Harrington,Taflove1,Taflove2}. For the hydrodynamic equations, there are 
also many well-designed numerical schemes. In addition to the Finite Difference (FD) method and the FE method, the Finite 
Volume (FV) method for conservation-type equations is very popular in Computational Fluid Dynamics (CFD), which is expected to 
achieve superior conservations, e.g. total energy conservation \cite{Anderson}. Because of their nature and the advantages 
in their respective fields, traditional numerical methods are often extended and reformed to simulate more complicated 
or hybrid systems \cite{Fang,QChen0}. When it comes to plasmonic phenomena, the simplest numerical treatment is introducing 
the DL model, which is a local linear response of the hydrodynamic equations into the FDTD or FE methods \cite{Barnes,
Ozbay}. This numerical model can be conveniently solved by circular convolution or Auxiliary Differential Equation (ADE) 
techniques \cite{Fang}. This linearized dispersion model provides us with abundant information about plasmonic perturbations, 
such as the linear dispersion relations and polarization modes, but the important nonlinear and dynamical properties, 
such as mode mixing, High Harmonic Generation (HHG), and the Kerr effect, are ignored \cite{Kim,Valev}. An FDTD-type method 
for Hydrodynamic-Maxwell equations of plasmonic metasurfaces can also be found, and many interesting results about the 
nonlinear effect are shown in the simulations \cite{Fang}. Previous work showed that plasmonic phenomena are attractive 
and important, which means that further numerical research is necessary. As a basic tool for computational plasmonics, more 
advanced numerical schemes for the free electron gas model are needed.

Because of the nonlinearity and the multi-scale nature of the Hydrodynamic-Maxwell equations, high-quality simulations 
of SPPs face challenges. For example, numerical errors involving the momentum and energy of the electron gas and the 
electromagnetic field can coherently accumulate, though these errors may be small in each numerical step. The breakdown 
of conservation laws over a long simulation time amounts to pseudo physics. It is desirable, therefore, to use numerical 
integrators with good global conservative properties. The canonical symplectic integrators for Hamiltonian systems with 
a canonical structure first developed by K. Feng \emph{et. al.} are known as a class of structure-preserving geometric 
algorithms that have excellent numerical performance in long-term simulations \cite{Feng1,Feng2,Benettin,Reich,Hairer1,
YWu,Hairer2,SChin,Qin1,MTao,Morrison1,QChen1}. Unfortunately, the hydrodynamic system does not possess a simple canonical 
structure, which means the canonical symplectic integrators do not apply directly \cite{Morrison2}. As an alternative 
method, J. Marsden and M. West developed the variational integrator based on the discrete Hamiltonian principle for 
Lagrangian systems \cite{Marsden,Lew,West}. The discrete variational principle preserves the Lagrangian structures of 
dynamical systems in Lagrangian form. As a significant advantage, the numerical errors of conserved quantities are bounded 
by a small value over a long time. The variational integrator has been widely used in many complex systems, especially in 
geophysics and plasma physics \cite{Qin2,Qin3,Qin4,Qin5,Qin6,Qin7,Stamm,Shadwick,Qin8,Morrison3,Qin9}.

In this work, we construct a class of variational schemes for geometric simulations of SPPs, and we show the advantages 
of the algorithms via several numerical experiments. In Sec.\ref{sec:2}, the Lagrangian density of lossless free electron 
gas with a self-consistent electromagnetic field is established, and the dynamics with constraints are obtained via the 
variational principle. With an appropriate derivation, the standard Hydrodynamic-Maxwell equations are obtained in an 
arbitrary gauge. In Sec.\ref{sec:3}, the action functional is discretized via Discrete Exterior Calculus (DEC) and minimized 
to obtain discrete dynamics. Equipped with efficient nonlinear and linear algebraic solvers, i.e., Newton-Raphson (N-R) 
iteration and the Biconjugate-Gradient-Stabilized (BICGSTAB) method, the discrete system is solvable. In App.\ref{sec:app}, 
the detailed Jacobian of nonlinear equations is derived, which guides the procedures of nonlinear iteration and associated 
linear updating. In Sec.\ref{sec:4}, the schemes are implemented to simulate bulk plasmons and SPPs. The numerical results 
reproduce the characteristic dispersion relations of the plasmonic systems, and they exhibit good long-term stability. 
These desirable features make the algorithm a powerful tool in the study of SPPs using the hydrodynamic-electrodynamic model.

\section{Model and Theory\label{sec:2}}

\subsection{Plasmonics\label{sec:2-1}}
\begin{figure}[htbp]
\centerline{\includegraphics[width=7.5cm,height=4.5cm]{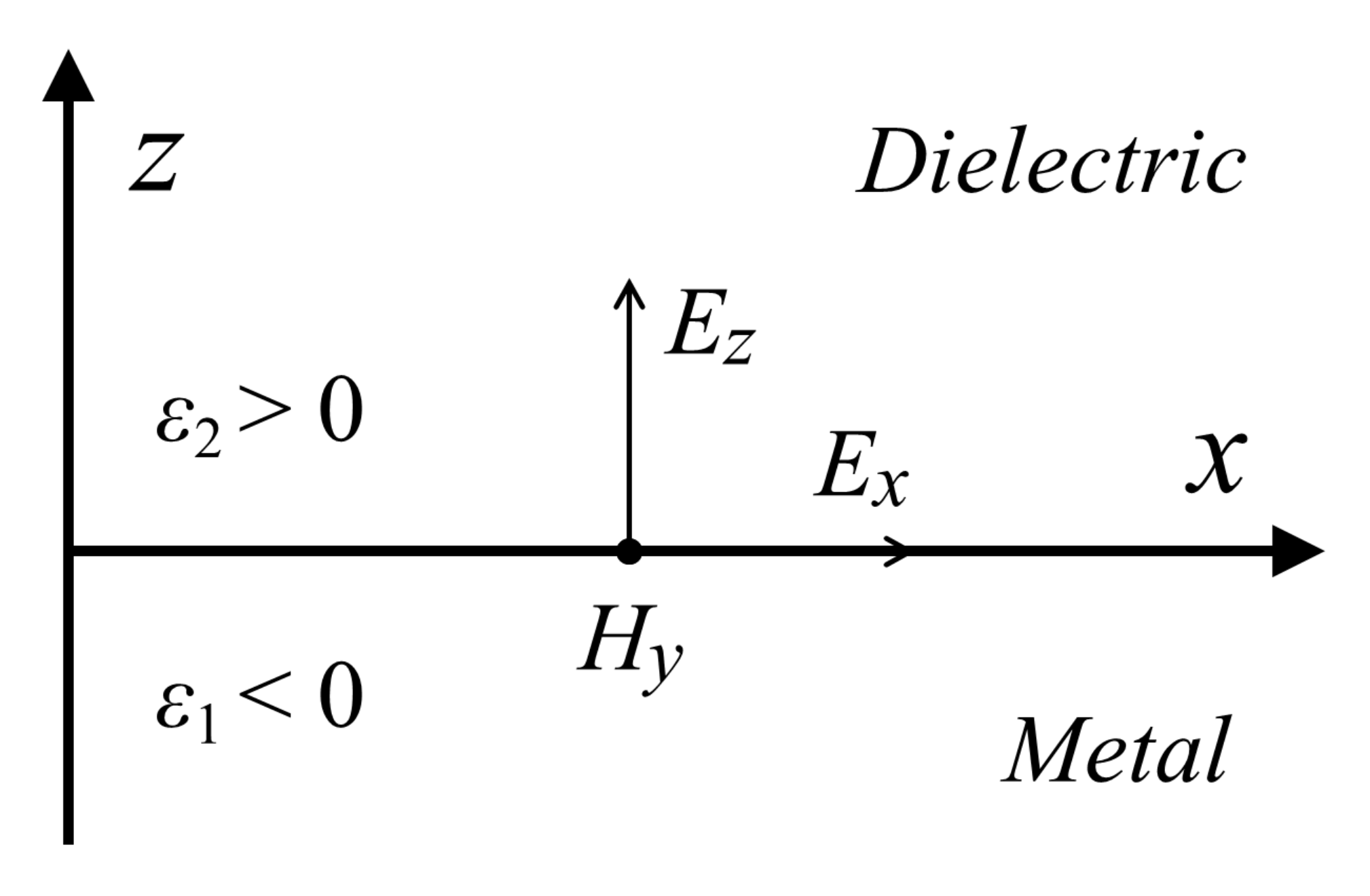}}
\caption{SPPs configuration at an infinite metal-dielectric interface.}
\label{fig:1}
\end{figure}

SPPs are a type of infrared or visible electromagnetic surface waves that travel along a metal-dielectric interface. 
The polariton means the wave consists of electron collective motion and a self-consistent electromagnetic field between 
the metal and the dielectric \cite{Barnes,Ozbay}. The simplest SPP configuration is shown in Fig.\ref{fig:1}, where a 
Transverse Magnetic (TM) mode propagates along an infinite interface between the metal (lower half-space) and the dielectric 
(upper half-space).

We assume that the media are nonmagnetic. At the region where the electromagnetic wave frequency $\omega$ is lower than 
the metal plasma frequency $\omega_{p}$, the metallic permittivity $\epsilon_{1}<0$ and the dielectric permittivity 
$\epsilon_{2}>0$. Then we can write the linearized SPPs field by using the Amp\`{e}re-Maxwell equation as,
\begin{eqnarray}
H_{y}\left(z\right)&=&\tilde{H}_{1}e^{ik_{x}x}e^{k_{z1}z},\label{eq:1}\\
E_{x}\left(z\right)&=&-i\tilde{H}_{1}\frac{k_{z1}}{\omega\epsilon_{0}\epsilon_{1}}e^{ik_{x}x}e^{k_{z1}z},\label{eq:2}\\
E_{z}\left(z\right)&=&-\tilde{H}_{1}\frac{k_{x}}{\omega\epsilon_{0}\epsilon_{1}}e^{ik_{x}x}e^{k_{z1}z},\label{eq:3}
\end{eqnarray}
for the metal region, and
\begin{eqnarray}
H_{y}\left(z\right)&=&\tilde{H}_{2}e^{ik_{x}x}e^{-k_{z2}z},\label{eq:4}\\
E_{x}\left(z\right)&=&i\tilde{H}_{2}\frac{k_{z2}}{\omega\epsilon_{0}\epsilon_{2}}e^{ik_{x}x}e^{-k_{z2}z},\label{eq:5}\\
E_{z}\left(z\right)&=&-\tilde{H}_{2}\frac{k_{x}}{\omega\epsilon_{0}\epsilon_{2}}e^{ik_{x}x}e^{-k_{z2}z},\label{eq:6}
\end{eqnarray}
for the dielectric region, where $\epsilon_{0}$ is the vacuum permittivity, and $\tilde{H}_{1}$ and $\tilde{H}_{2}$ are 
the magnetic field amplitudes. Applying the interface conditions $E_{x1}(0)=E_{x2}(0)$ and $H_{y1}(0)=H_{y2}(0)$ into 
Eqs.~\eqref{eq:1}-\eqref{eq:6}, we can obtain $\tilde{H}_{1}=\tilde{H}_{2}$ and $\epsilon_{1}k_{z2}=-\epsilon_{2}k_{z1}$. 
Then the linearized SPPs dispersion relations are given by using Faraday's equation as \cite{Barnes},
\begin{eqnarray}
k^{2}_{z1}&=&k^{2}_{x}-k^{2}_{0}\epsilon_{1},\label{eq:7}\\
k^{2}_{z2}&=&k^{2}_{x}-k^{2}_{0}\epsilon_{2},\label{eq:8}\\
k_{x}&=&k_{0}\sqrt{\frac{\epsilon_{1}\epsilon_{2}}{\epsilon_{1}+\epsilon_{2}}},\label{eq:9}
\end{eqnarray}
where $k_{0}=\omega/c$ is the vacuum wave number.

The previous descriptions for SPPs provide us with a basic physical image and important spectral information. 
At the low-frequency (terahertz or millimeter) region, Eq.~\eqref{eq:9} reduces to $k_{x}\approx{k_{0}}$ and the 
dielectric confinement strength $k_{z2}$ has a very small value, which means that the SPPs reduce to weak-bound 
Sommerfeld-Zenneck (SZ) waves. At the high-frequency (visible and near infrared) region, Eq.~\eqref{eq:9} reduces to 
$\omega\approx{\omega_{sp}}=\omega_{p}/\sqrt{1+\epsilon_{2}}$, where $\omega_{sp}$ is the Surface Plasmon Resonance 
(SPR) frequency. This means that the SPPs at the resonance region are far from the light cone, which has infinitesimal 
group velocity $v_{g}$ and phase velocity $v_{p}$. Both the dielectric and metal confinement strengths $k_{z2,1}$ 
of these quasi-static modes near the resonance region have very large values, which means the SPPs energy is strongly 
localized. As the cornerstone of many plasmonic techniques, the strong-bound sub-wave-length field is an important 
feature of the SPPs.

\subsection{Lagrangian Form of Lossless Free Electron Gas\label{sec:2-2}}
The Lagrangian density of lossless free electron gas in a quasi-neutral background with a self-consistent electromagnetic 
field can be defined from the hydrodynamic form as \cite{Newcomb,Sahraoui},
\begin{eqnarray}
\mathcal{L}&=&\mathcal{L}_{\rm{EG}}+\mathcal{L}_{\rm{EM}}+\mathcal{L}_{\rm{Int}},\label{eq:10}
\end{eqnarray}
\begin{eqnarray}
\mathcal{L}_{\rm{EG}}&=&\frac{1}{2}mn\bm{v}^2+\alpha\left[\frac{\partial}{\partial{t}}n+\bigtriangledown\cdot\left(n\bm{v}\right)\right]-\lambda\left(\frac{\partial}{\partial{t}}\mu+\bm{v}\cdot\bigtriangledown\mu\right),\label{eq:ldeg}
\end{eqnarray}
\begin{eqnarray}
\mathcal{L}_{\rm{EM}}&=&\frac{\epsilon_{0}}{2}\left(-\bigtriangledown\phi-\frac{\partial}{\partial{t}}\bm{A}\right)^2-\frac{1}{2\mu_{0}}(\bigtriangledown\times\bm{A})^2,\label{eq:ldem}
\end{eqnarray}
\begin{eqnarray}
\mathcal{L}_{\rm{Int}}&=&en\bm{v}\cdot\bm{A}-e\left(n-n_{0}\right)\phi,\label{eq:ldint}
\end{eqnarray}
where the subscripts denote Electron Gas (EG), ElectroMagnetic (EM), and Interaction (Int), respectively. $n$ is the 
electron density, $n_{0}$ is the background particle density, $\bm{v}$ is the electron gas velocity, $e$ is the 
electron charge and $m$ is the electron mass. The electromagnetic field components $(\phi, \bm{A})$ are defined in 
an arbitrary gauge. This Lagrangian density is given in constraint form, where the scalar fields $\alpha$ and $\lambda$ 
are Lagrangian multipliers. In the last term of Eq.~\eqref{eq:10}, Lin's constraint factor $\mu$ is used to establish 
a complete description for the velocity and helicity \cite{Newcomb,Sahraoui}.

With the Lagrangian density \eqref{eq:10}, we can construct the action functional 
$S=\int_{T}\int_{\Omega}\mathcal{L}{\rm{d}}x^3{\rm{d}}t$ which involves the physics of lossless free electron gas. 
With tedious variational calculation, the total variation of $S$ with fixed boundary can be given as,
\begin{eqnarray}
\delta{S}&=&\int_{T}\int_{\Omega}\left\{\left(\frac{1}{2}m\bm{v}^{2}+e\bm{v}\cdot\bm{A}-e\phi-\frac{\partial}{\partial{t}}\alpha-\bm{v}\cdot\bigtriangledown\alpha\right)\delta{n}\right.\nonumber\\
& &+\left(mn\bm{v}+en\bm{A}-n\bigtriangledown\alpha-\lambda\bigtriangledown\mu\right)\cdot\delta\bm{v}\nonumber\\
& &+\left[en\bm{v}-\epsilon_{0}\left(\bigtriangledown\frac{\partial}{\partial{t}}\phi+\frac{\partial^{2}}{\partial{t}^{2}}\bm{A}\right)-\frac{1}{\mu_{0}}\bigtriangledown\times\bigtriangledown\times\bm{A}\right]\cdot\delta\bm{A}\nonumber\\
& &-\left[e\left(n-n_{0}\right)-\epsilon_{0}\left(\bigtriangledown^{2}\phi+\bigtriangledown\cdot\frac{\partial}{\partial{t}}\bm{A}\right)\right]\delta\phi\nonumber\\
& &+\left[\frac{\partial}{\partial{t}}n+\bigtriangledown\cdot\left(n\bm{v}\right)\right]\delta\alpha-\left(\frac{\partial}{\partial{t}}\mu+\bm{v}\cdot\bigtriangledown\mu\right)\delta\lambda\nonumber\\
& &\left.+\left[\frac{\partial}{\partial{t}}\lambda+\bigtriangledown\cdot\left(\lambda\bm{v}\right)\right]\delta\mu\right\}{\rm{d}}x^3{\rm{d}}t.\label{eq:11}
\end{eqnarray}

Taking variational derivative of the action functional $S$ with respect to the fields 
$q=(n,\bm{v},\bm{A},\phi,\alpha,\lambda,\mu)$ and minimizing $S$ by leading variational derivatives equal to 0, we can 
obtain the Euler-Lagrange equations of lossless free electron gas,
\begin{eqnarray}
\frac{\delta{S}}{\delta{n}}=0\Rightarrow\nonumber\\
\frac{\partial}{\partial{t}}\alpha=\frac{1}{2}m\bm{v}^{2}-\bm{v}\cdot\bigtriangledown\alpha+e\bm{v}\cdot\bm{A}-e\phi,\label{eq:12}\\
\frac{\delta{S}}{\delta{\bm{v}}}=0\Rightarrow\nonumber\\
mn\bm{v}+en\bm{A}=n\bigtriangledown\alpha+\lambda\bigtriangledown\mu,\label{eq:13}\\
\frac{\delta{S}}{\delta{\bm{A}}}=0\Rightarrow\nonumber\\
\frac{\partial^{2}}{\partial{t}^{2}}\bm{A}=-\frac{1}{\epsilon_{0}\mu_{0}}\bigtriangledown\times\bigtriangledown\times\bm{A}-\bigtriangledown\frac{\partial}{\partial{t}}\phi+\frac{e}{\epsilon_{0}}n\bm{v},\label{eq:14}\\
\frac{\delta{S}}{\delta\phi}=0\Rightarrow\nonumber\\
\bigtriangledown^{2}\phi+\bigtriangledown\cdot\frac{\partial}{\partial{t}}\bm{A}=-\frac{e}{\epsilon_{0}}\left(n-n_{0}\right),\label{eq:15}\\
\frac{\delta{S}}{\delta\alpha}=0\Rightarrow\nonumber\\
\frac{\partial}{\partial{t}}n=-\bigtriangledown\cdot\left(n\bm{v}\right),\label{eq:16}\\
\frac{\delta{S}}{\delta\lambda}=0\Rightarrow\nonumber\\
\frac{\partial}{\partial{t}}\mu=-\bm{v}\cdot\bigtriangledown\mu,\label{eq:17}\\
\frac{\delta{S}}{\delta\mu}=0\Rightarrow\nonumber\\
\frac{\partial}{\partial{t}}\lambda=-\bigtriangledown\cdot\left(\lambda\bm{v}\right).\label{eq:18}
\end{eqnarray}
Eqs.~\eqref{eq:12} and \eqref{eq:18} are Lagrangian multiplier equations. Eq.~\eqref{eq:13} defines the canonical 
momentum density of free electron gas. Eqs.~\eqref{eq:14} and \eqref{eq:15} are Maxwell's equations. Eq.~\eqref{eq:16} 
is the continuity equation of free electron gas. Eq.~\eqref{eq:17} defines the dynamics of Lin's constraint field. 
These equations provide us with a complete model to describe plasmonic phenomena.

\subsection{Hydrodynamic-Maxwell Model\label{sec:2-3}}
The Euler-Lagrange equations \eqref{eq:12}-\eqref{eq:18} can be recognized as a general form of the well-known 
hydrodynamic-Maxwell equations. Here we give a detailed derivation to illustrate the relation between two kinds 
of equations.

Adopting the temporal gauge $\phi=0$ explicitly, we can define the electromagnetic field components as 
$\bm{E}=-\partial\bm{A}/\partial{t}$ and $\mu_{0}\bm{H}=\bigtriangledown\times\bm{A}$. Taking the derivative of 
Eq.~\eqref{eq:13} with respect to $t$, we obtain,
\begin{eqnarray}
m\left(\frac{\partial}{\partial{t}}n\right)\bm{v}+mn\frac{\partial}{\partial{t}}\bm{v}+e\left(\frac{\partial}{\partial{t}}n\right)\bm{A}+en\frac{\partial}{\partial{t}}\bm{A}\nonumber\\
=\left(\frac{\partial}{\partial{t}}n\right)\bigtriangledown\alpha+n\bigtriangledown\frac{\partial}{\partial{t}}\alpha+\left(\frac{\partial}{\partial{t}}\lambda\right)\bigtriangledown\mu+\lambda\bigtriangledown\frac{\partial}{\partial{t}}\mu.\label{eq:19}
\end{eqnarray}
With tedious algebraic calculation, we obtain the momentum equation of free electron gas by substituting 
Eqs.~\eqref{eq:12} and \eqref{eq:16}-\eqref{eq:18} into Eq.~\eqref{eq:19} as,
\begin{eqnarray}
mn\frac{\partial}{\partial{t}}\bm{v}&=&\left(mn\bm{v}+en\bm{A}-n\bigtriangledown\alpha-\lambda\bigtriangledown\mu\right)\cdot\bigtriangledown\bm{v}\nonumber\\
& &+\left(mn\bm{v}+en\bm{A}-n\bigtriangledown\alpha-\lambda\bigtriangledown\mu\right)\times\bigtriangledown\times\bm{v}\nonumber\\
& &+\bm{v}\cdot\left(\frac{\lambda}{n}\bigtriangledown{n}\bigtriangledown\mu-\bigtriangledown\lambda\bigtriangledown\mu-\lambda\bigtriangledown\bigtriangledown\mu-n\bigtriangledown\bigtriangledown\alpha+en\bigtriangledown\bm{A}\right)\nonumber\\
& &+en\left[-\frac{\partial}{\partial{t}}\bm{A}+\bm{v}\times\left(\bigtriangledown\times\bm{A}\right)\right].\label{eq:20}
\end{eqnarray}
Then taking the gradient $\bigtriangledown$ on Eq.~\eqref{eq:13}, we obtain,
\begin{eqnarray}
-mn\bigtriangledown\bm{v}&=&en\bigtriangledown\bm{A}-\bigtriangledown\lambda\bigtriangledown\mu+\frac{\lambda}{n}\bigtriangledown{n}\bigtriangledown\mu-\lambda\bigtriangledown\bigtriangledown\mu-n\bigtriangledown\bigtriangledown\alpha.\label{eq:21}
\end{eqnarray}
At last, we obtain the velocity equation of free electron gas by substituting Eqs.~\eqref{eq:13} and \eqref{eq:21} 
into Eq.~\eqref{eq:20} as,
\begin{eqnarray}
\frac{\partial}{\partial{t}}\bm{v}=-\bm{v}\cdot\bigtriangledown\bm{v}+\frac{e}{m}\left[-\frac{\partial}{\partial{t}}\bm{A}+\bm{v}\times\left(\bigtriangledown\times\bm{A}\right)\right].\label{eq:22}
\end{eqnarray}

Based on the previous derivation, Eqs.~\eqref{eq:14}-\eqref{eq:16} and \eqref{eq:22} make up the standard 
hydrodynamic-Maxwell equations, which can be rewritten in a compact form as,
\begin{eqnarray}
\frac{\partial}{\partial{t}}n+\bigtriangledown\cdot\left(n\bm{v}\right)=0,\label{eq:23}\\
\frac{\partial}{\partial{t}}\bm{v}+\bm{v}\cdot\bigtriangledown\bm{v}=\frac{e}{m}\left(\bm{E}+\mu_{0}\bm{v}\times\bm{H}\right),\label{eq:24}\\
\frac{1}{c^{2}}\frac{\partial^{2}}{\partial{t}^{2}}\bm{A}+\bigtriangledown\times\bigtriangledown\times\bm{A}=\mu_{0}en\bm{v},\label{eq:25}\\
\bigtriangledown\cdot\frac{\partial}{\partial{t}}\bm{A}=-\frac{e}{\epsilon_{0}}\left(n-n_{0}\right).\label{eq:26}
\end{eqnarray}

The hydrodynamic-Maxwell equations \eqref{eq:23}-\eqref{eq:26} are complete and self-consistent. The electromagnetic 
field components enter the hydrodynamic velocity equation \eqref{eq:24} via Lorentz force. The hydrodynamic fields 
enter Maxwell's equations \eqref{eq:25}-\eqref{eq:26} via the current and charge densities. In these equations, both 
the dynamics of electron collective motion and self-consistent field are involved. Based on previous discussion, both 
the hydrodynamic-Maxwell equations \eqref{eq:23}-\eqref{eq:26} and Euler-Lagrange equations \eqref{eq:12}-\eqref{eq:18} 
of lossless free electron gas can be used as the basic physical model of plasmonics.

\section{Numerical Strategies\label{sec:3}}

\subsection{DEC Based Discretization\label{sec:3-1}}
The traditional numerical methods for an infinite dimensional dynamical system focus on the solving techniques for 
relevant differential or integral equations, such as the FDTD, FEM or FV types of numerical schemes for the 
hydrodynamic-Maxwell equations \eqref{eq:23}-\eqref{eq:26} or even the alternative Eqs.~\eqref{eq:12}-\eqref{eq:18} 
in plasmonic research. Here we construct a numerical strategy for hydrodynamic-electrodynamic model based plasmonic 
phenomena simulations in a different way. Instead of directly discretizing the previous differential equations, we 
reconstruct the discrete dynamics via discrete variational principle, which is an infinite dimensional Hamilton's 
principle analog on discrete space-time manifold. These variational schemes have particular advantages in high-quality 
long term simulations, as they can preserve the discrete Lagrangian symplectic structure, gauge symmetry and general 
energy-momentum density, which will be discussed in detail in the following sections.

\begin{figure}[htbp]
\centerline{\includegraphics[width=8cm,height=6cm]{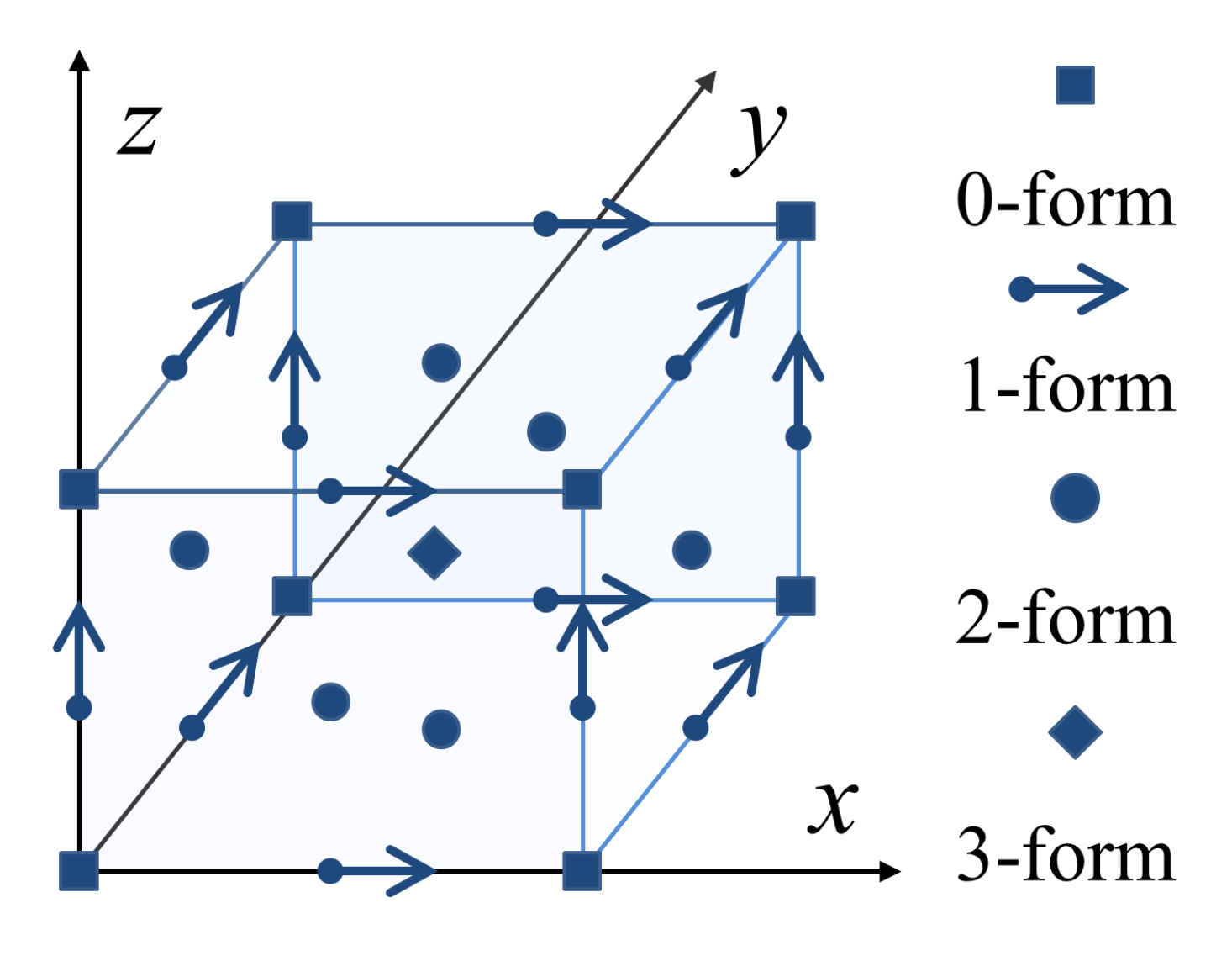}}
\caption{DEC-based discretization of the hydrodynamic-electrodynamic model in a rectangular lattice. The discrete 
spacelike submanifold is shown, and other types of discrete submanifolds can be given in the same way.}
\label{fig:2}
\end{figure}

The first step in numerical simulation is discretization. As a differential geometry based numerical framework, DEC 
defines complete operational rules on a discrete differential manifold, which form a cochain complex \cite{Morrison3,
Hirani,Hiptmair,Arnold1,Arnold2,Holst}. To solve the continuous system numerically, the space-time manifold is 
discretized using a rectangular lattice (other lattices are also viable). Then the scalar fields, which are 3-forms, 
i.e. $ndx{\wedge}dy{\wedge}dz$ and ${\lambda}dx{\wedge}dy{\wedge}dz$, naturally live on the volume center of the discrete 
spacelike submanifold,
\begin{eqnarray}
n^{n}_{i+\frac{1}{2},j+\frac{1}{2},k+\frac{1}{2}}:n\left(t_{n},x_{i}+\frac{\Delta{x}}{2},y_{j}+\frac{\Delta{y}}{2},z_{k}+\frac{\Delta{z}}{2}\right),\label{eq:27}
\end{eqnarray}
where $(t_{n},x_{i},y_{j},z_{k})$ is the coordinate of the lattice vertex. The velocity, multiplier, constraint and 
gauge 1-forms $\bm{v}=v_{i}dx^{i}$, $\alpha{dt}$, $\mu{dt}$ and $\bm{A}=A_{\nu}dx^{\nu}$ naturally live along the edges 
of the space-time lattice,
\begin{eqnarray}
\phi^{n+\frac{1}{2}}_{i,j,k}:\phi\left(t_{n}+\frac{\Delta{t}}{2},x_{i},y_{j},z_{k}\right),\label{eq:28}\\
A^{n}_{xi+\frac{1}{2},j,k}:A_{x}\left(t_{n},x_{i}+\frac{\Delta{x}}{2},y_{j},z_{k}\right),\label{eq:29}\\
A^{n}_{yi,j+\frac{1}{2},k}:A_{y}\left(t_{n},x_{i},y_{j}+\frac{\Delta{y}}{2},z_{k}\right),\label{eq:30}\\
A^{n}_{zi,j,k+\frac{1}{2}}:A_{z}\left(t_{n},x_{i},y_{j},z_{k}+\frac{\Delta{z}}{2}\right).\label{eq:31}
\end{eqnarray}
In the above discretization, a half integer index indicates along which edge does the field resides. Then the 
2-forms, e.g., $\bm{F}=\bm{{\rm{d}}}\bm{A}$, and the 4-forms, e.g., $\bm{F}\wedge*\bm{F}$, are defined on the faces 
and volume center of the space-time lattice respectively, where $\bm{{\rm{d}}}$ is the exterior derivative operator. 
The discretization of the spacelike submanifold is shown in Fig.\ref{fig:2}.

Based on these definitions, the exterior derivatives of discrete differential forms are naturally obtained via the 
forward difference operators. The exterior derivatives on the spacelike submanifold are given as,
\begin{eqnarray}
\bm{{\rm{d}}}\phi=\left(\bigtriangledown\phi\right)_{i}dx^{i}=\frac{\phi_{i+1,j,k}-\phi_{i,j,k}}{\Delta{x}}dx+\frac{\phi_{i,j+1,k}-\phi_{i,j,k}}{\Delta{y}}dy+\frac{\phi_{i,j,k+1}-\phi_{i,j,k}}{\Delta{z}}dz,\label{eq:32}
\end{eqnarray}
\begin{eqnarray}
\bm{{\rm{d}}}\bm{A}&=&\left(\bigtriangledown\times\bm{A}\right)_{i}dx^{j}{\wedge}dx^{k}\nonumber\\
&=&\left(\frac{A_{zi,j+1,k+\frac{1}{2}}-A_{zi,j,k+\frac{1}{2}}}{\Delta{y}}-\frac{A_{yi,j+\frac{1}{2},k+1}-A_{yi,j+\frac{1}{2},k}}{\Delta{z}}\right)dy{\wedge}dz\nonumber\\
& &+\left(\frac{A_{xi+\frac{1}{2},j,k+1}-A_{xi+\frac{1}{2},j,k}}{\Delta{z}}-\frac{A_{zi+1,j,k+\frac{1}{2}}-A_{zi,j,k+\frac{1}{2}}}{\Delta{x}}\right)dz{\wedge}dx\nonumber\\
& &+\left(\frac{A_{yi+1,j+\frac{1}{2},k}-A_{yi,j+\frac{1}{2},k}}{\Delta{x}}-\frac{A_{xi+\frac{1}{2},j+1,k}-A_{xi+\frac{1}{2},j,k}}{\Delta{y}}\right)dx{\wedge}dy,\label{eq:33}
\end{eqnarray}
\begin{eqnarray}
\bm{{\rm{d}}}*\bm{A}=\bigtriangledown\cdot\bm{A}dx^{i}{\wedge}dx^{j}{\wedge}dx^{k}\nonumber\\
=\left(\frac{A_{xi+\frac{1}{2},j,k}-A_{xi-\frac{1}{2},j,k}}{\Delta{x}}+\frac{A_{yi,j+\frac{1}{2},k}-A_{yi,j-\frac{1}{2},k}}{\Delta{y}}+\frac{A_{zi,j,k+\frac{1}{2}}-A_{zi,j,k-\frac{1}{2}}}{\Delta{z}}\right)dx{\wedge}dy{\wedge}dz,\label{eq:34}
\end{eqnarray}
where $*$ is the Hodge star operator, which generates the Hodge dual form of the primary discrete form.

By using the DEC, the Lagrangian density \eqref{eq:10} in the interval $[t_{n},t_{n+1}]$ can be discretized as,
\begin{eqnarray}
\mathcal{L}^{n{\sim}n+1}_{di,j,k}&=&\mathcal{L}^{n{\sim}n+1}_{d{\rm{EG}}i,j,k}+\mathcal{L}^{n{\sim}n+1}_{d{\rm{EM}}i,j,k}+\mathcal{L}^{n{\sim}n+1}_{d{\rm{Int}}i,j,k},\label{eq:35}
\end{eqnarray}
\begin{eqnarray}
\mathcal{L}^{n{\sim}n+1}_{d{\rm{EG}}i,j,k}&=&\frac{1}{2}mn^{n}_{i+\frac{1}{2},j+\frac{1}{2},k+\frac{1}{2}}\left(v^{n^2}_{xi+\frac{1}{2},j,k}+v^{n^2}_{yi,j+\frac{1}{2},k}+v^{n^2}_{zi,j,k+\frac{1}{2}}\right)\nonumber\\
& &+\alpha^{n+\frac{1}{2}}_{i,j,k}\left(\frac{n^{n+1}_{i+\frac{1}{2},j+\frac{1}{2},k+\frac{1}{2}}-n^{n}_{i+\frac{1}{2},j+\frac{1}{2},k+\frac{1}{2}}}{\Delta{t}}\right.\nonumber\\
& &+\frac{n^{n+1}_{i+\frac{1}{2},j+\frac{1}{2},k+\frac{1}{2}}v^{n+1}_{xi+\frac{1}{2},j,k}-n^{n+1}_{i-\frac{1}{2},j+\frac{1}{2},k+\frac{1}{2}}v^{n+1}_{xi-\frac{1}{2},j,k}}{\Delta{x}}\nonumber\\
& &+\frac{n^{n+1}_{i+\frac{1}{2},j+\frac{1}{2},k+\frac{1}{2}}v^{n+1}_{yi,j+\frac{1}{2},k}-n^{n+1}_{i+\frac{1}{2},j-\frac{1}{2},k+\frac{1}{2}}v^{n+1}_{yi,j-\frac{1}{2},k}}{\Delta{y}}\nonumber\\
& &\left.+\frac{n^{n+1}_{i+\frac{1}{2},j+\frac{1}{2},k+\frac{1}{2}}v^{n+1}_{zi,j,k+\frac{1}{2}}-n^{n+1}_{i+\frac{1}{2},j+\frac{1}{2},k-\frac{1}{2}}v^{n+1}_{zi,j,k-\frac{1}{2}}}{\Delta{z}}\right)\nonumber\\
& &-\lambda^{n}_{i+\frac{1}{2},j+\frac{1}{2},k+\frac{1}{2}}\left(\frac{\mu^{n+\frac{1}{2}}_{i,j,k}-\mu^{n-\frac{1}{2}}_{i,j,k}}{\Delta{t}}+v^{n}_{xi+\frac{1}{2},j,k}\frac{\mu^{n-\frac{1}{2}}_{i+1,j,k}-\mu^{n-\frac{1}{2}}_{i,j,k}}{\Delta{x}}\right.\nonumber\\
& &\left.+v^{n}_{yi,j+\frac{1}{2},k}\frac{\mu^{n-\frac{1}{2}}_{i,j+1,k}-\mu^{n-\frac{1}{2}}_{i,j,k}}{\Delta{y}}+v^{n}_{zi,j,k+\frac{1}{2}}\frac{\mu^{n-\frac{1}{2}}_{i,j,k+1}-\mu^{n-\frac{1}{2}}_{i,j,k}}{\Delta{z}}\right),\label{eq:dleg}
\end{eqnarray}
\begin{eqnarray}
\mathcal{L}^{n{\sim}n+1}_{d{\rm{EM}}i,j,k}&=&\frac{\epsilon_{0}}{2}\left[\left(\frac{\phi^{n+\frac{1}{2}}_{i+1,j,k}-\phi^{n+\frac{1}{2}}_{i,j,k}}{\Delta{x}}+\frac{A^{n+1}_{xi+\frac{1}{2},j,k}-A^{n}_{xi+\frac{1}{2},j,k}}{\Delta{t}}\right)^{2}\right.\nonumber\\
& &+\left(\frac{\phi^{n+\frac{1}{2}}_{i,j+1,k}-\phi^{n+\frac{1}{2}}_{i,j,k}}{\Delta{y}}+\frac{A^{n+1}_{yi,j+\frac{1}{2},k}-A^{n}_{yi,j+\frac{1}{2},k}}{\Delta{t}}\right)^{2}\nonumber\\
& &\left.+\left(\frac{\phi^{n+\frac{1}{2}}_{i,j,k+1}-\phi^{n+\frac{1}{2}}_{i,j,k}}{\Delta{z}}+\frac{A^{n+1}_{zi,j,k+\frac{1}{2}}-A^{n}_{zi,j,k+\frac{1}{2}}}{\Delta{t}}\right)^{2}\right]\nonumber\\
& &-\frac{1}{2\mu_{0}}\left[\left(\frac{A^{n}_{zi,j+1,k+\frac{1}{2}}-A^{n}_{zi,j,k+\frac{1}{2}}}{\Delta{y}}-\frac{A^{n}_{yi,j+\frac{1}{2},k+1}-A^{n}_{yi,j+\frac{1}{2},k}}{\Delta{z}}\right)^{2}\right.\nonumber\\
& &+\left(\frac{A^{n}_{xi+\frac{1}{2},j,k+1}-A^{n}_{xi+\frac{1}{2},j,k}}{\Delta{z}}-\frac{A^{n}_{zi+1,j,k+\frac{1}{2}}-A^{n}_{zi,j,k+\frac{1}{2}}}{\Delta{x}}\right)^{2}\nonumber\\
& &\left.+\left(\frac{A^{n}_{yi+1,j+\frac{1}{2},k}-A^{n}_{yi,j+\frac{1}{2},k}}{\Delta{x}}-\frac{A^{n}_{xi+\frac{1}{2},j+1,k}-A^{n}_{xi+\frac{1}{2},j,k}}{\Delta{y}}\right)^{2}\right],\label{eq:dlem}
\end{eqnarray}
\begin{eqnarray}
\mathcal{L}^{n{\sim}n+1}_{d{\rm{Int}}i,j,k}&=&en^{n}_{i+\frac{1}{2},j+\frac{1}{2},k+\frac{1}{2}}\left(v^{n}_{xi+\frac{1}{2},j,k}A^{n}_{xi+\frac{1}{2},j,k}+v^{n}_{yi,j+\frac{1}{2},k}A^{n}_{yi,j+\frac{1}{2},k}+v^{n}_{zi,j,k+\frac{1}{2}}A^{n}_{zi,j,k+\frac{1}{2}}\right)\nonumber\\
& &-e\left(n^{n}_{i+\frac{1}{2},j+\frac{1}{2},k+\frac{1}{2}}-n_{0}\right)\phi^{n+\frac{1}{2}}_{i,j,k}.\label{eq:dlint}
\end{eqnarray}
Then the action functional of discrete dynamical system is,
\begin{eqnarray}
S_{d}=\sum^{N}_{n=0}L^{n{\sim}n+1}_{d}\Delta{t},\label{eq:36}
\end{eqnarray}
\begin{eqnarray}
L^{n{\sim}n+1}_{d}=\sum_{i,j,k}\mathcal{L}^{n{\sim}n+1}_{di,j,k}\Delta{x}\Delta{y}\Delta{z},\label{eq:37}
\end{eqnarray}
where the functional $L^{n{\sim}n+1}_{d}$ is the Lagrangian of discrete dynamical system in interval $[t_{n},t_{n+1}]$.

\subsection{Variational Schemes\label{sec:3-2}}
With DEC-based discretization, we reconstruct a field theory on the discrete space-time manifold. Then the variational 
derivatives of action $S_{d}$ with respect to fields reduce to partial derivatives with respect to discrete differential 
forms. By minimizing the action, we obtain the discrete dynamical equations,
\begin{eqnarray}
\frac{\partial{S_{d}}}{\partial{n^{n}_{i+\frac{1}{2},j+\frac{1}{2},k+\frac{1}{2}}}}=\frac{\partial(L^{n-1{\sim}n}_{d}+L^{n{\sim}n+1}_{d})}{\partial{n^{n}_{i+\frac{1}{2},j+\frac{1}{2},k+\frac{1}{2}}}}\Delta{t}=0\Rightarrow\nonumber\\
\alpha^{n+\frac{1}{2}}_{i,j,k}-\frac{m\Delta{t}}{2}\left(v^{n^{2}}_{xi+\frac{1}{2},j,k}+v^{n^{2}}_{yi,j+\frac{1}{2},k}+v^{n^{2}}_{zi,j,k+\frac{1}{2}}\right)\nonumber\\
-e\Delta{t}\left(v^{n}_{xi+\frac{1}{2},j,k}A^{n}_{xi+\frac{1}{2},j,k}+v^{n}_{yi,j+\frac{1}{2},k}A^{n}_{yi,j+\frac{1}{2},k}+v^{n}_{zi,j,k+\frac{1}{2}}A^{n}_{zi,j,k+\frac{1}{2}}\right)+e\Delta{t}\phi^{n+\frac{1}{2}}_{i,j,k}\nonumber\\
+\frac{\Delta{t}}{\Delta{x}}v^{n}_{xi+\frac{1}{2},j,k}\left(\alpha^{n-\frac{1}{2}}_{i+1,j,k}-\alpha^{n-\frac{1}{2}}_{i,j,k}\right)+\frac{\Delta{t}}{\Delta{y}}v^{n}_{yi,j+\frac{1}{2},k}\left(\alpha^{n-\frac{1}{2}}_{i,j+1,k}-\alpha^{n-\frac{1}{2}}_{i,j,k}\right)\nonumber\\
+\frac{\Delta{t}}{\Delta{z}}v^{n}_{zi,j,k+\frac{1}{2}}\left(\alpha^{n-\frac{1}{2}}_{i,j,k+1}-\alpha^{n-\frac{1}{2}}_{i,j,k}\right)=\alpha^{n-\frac{1}{2}}_{i,j,k},\label{eq:38}
\end{eqnarray}
\begin{eqnarray}
\frac{\partial{S_{d}}}{\partial{v^{n}_{xi+\frac{1}{2},j,k}}}=\frac{\partial(L^{n-1{\sim}n}_{d}+L^{n{\sim}n+1}_{d})}{\partial{v^{n}_{xi+\frac{1}{2},j,k}}}\Delta{t}=0\Rightarrow\nonumber\\
mn^{n}_{i+\frac{1}{2},j+\frac{1}{2},k+\frac{1}{2}}v^{n}_{xi+\frac{1}{2},j,k}+en^{n}_{i+\frac{1}{2},j+\frac{1}{2},k+\frac{1}{2}}A^{n}_{xi+\frac{1}{2},j,k}\nonumber\\
=n^{n}_{i+\frac{1}{2},j+\frac{1}{2},k+\frac{1}{2}}\frac{\alpha^{n-\frac{1}{2}}_{i+1,j,k}-\alpha^{n-\frac{1}{2}}_{i,j,k}}{\Delta{x}}+\lambda^{n}_{i+\frac{1}{2},j+\frac{1}{2},k+\frac{1}{2}}\frac{\mu^{n-\frac{1}{2}}_{i+1,j,k}-\mu^{n-\frac{1}{2}}_{i,j,k}}{\Delta{x}},\label{eq:39}
\end{eqnarray}
\begin{eqnarray}
\frac{\partial{S_{d}}}{\partial{v^{n}_{yi,j+\frac{1}{2},k}}}=\frac{\partial(L^{n-1{\sim}n}_{d}+L^{n{\sim}n+1}_{d})}{\partial{v^{n}_{yi,j+\frac{1}{2},k}}}\Delta{t}=0\Rightarrow\nonumber\\
mn^{n}_{i+\frac{1}{2},j+\frac{1}{2},k+\frac{1}{2}}v^{n}_{yi,j+\frac{1}{2},k}+en^{n}_{i+\frac{1}{2},j+\frac{1}{2},k+\frac{1}{2}}A^{n}_{yi,j+\frac{1}{2},k}\nonumber\\
=n^{n}_{i+\frac{1}{2},j+\frac{1}{2},k+\frac{1}{2}}\frac{\alpha^{n-\frac{1}{2}}_{i,j+1,k}-\alpha^{n-\frac{1}{2}}_{i,j,k}}{\Delta{y}}+\lambda^{n}_{i+\frac{1}{2},j+\frac{1}{2},k+\frac{1}{2}}\frac{\mu^{n-\frac{1}{2}}_{i,j+1,k}-\mu^{n-\frac{1}{2}}_{i,j,k}}{\Delta{y}},\label{eq:40}
\end{eqnarray}
\begin{eqnarray}
\frac{\partial{S_{d}}}{\partial{v^{n}_{zi,j,k+\frac{1}{2}}}}=\frac{\partial(L^{n-1{\sim}n}_{d}+L^{n{\sim}n+1}_{d})}{\partial{v^{n}_{zi,j,k+\frac{1}{2}}}}\Delta{t}=0\Rightarrow\nonumber\\
mn^{n}_{i+\frac{1}{2},j+\frac{1}{2},k+\frac{1}{2}}v^{n}_{zi,j,k+\frac{1}{2}}+en^{n}_{i+\frac{1}{2},j+\frac{1}{2},k+\frac{1}{2}}A^{n}_{zi,j,k+\frac{1}{2}}\nonumber\\
=n^{n}_{i+\frac{1}{2},j+\frac{1}{2},k+\frac{1}{2}}\frac{\alpha^{n-\frac{1}{2}}_{i,j,k+1}-\alpha^{n-\frac{1}{2}}_{i,j,k}}{\Delta{z}}+\lambda^{n}_{i+\frac{1}{2},j+\frac{1}{2},k+\frac{1}{2}}\frac{\mu^{n-\frac{1}{2}}_{i,j,k+1}-\mu^{n-\frac{1}{2}}_{i,j,k}}{\Delta{z}},\label{eq:41}
\end{eqnarray}
\begin{eqnarray}
\frac{\partial{S_{d}}}{\partial{A^{n}_{xi+\frac{1}{2},j,k}}}=\frac{\partial(L^{n-1{\sim}n}_{d}+L^{n{\sim}n+1}_{d})}{\partial{A^{n}_{xi+\frac{1}{2},j,k}}}\Delta{t}=0\Rightarrow\nonumber\\
\frac{\epsilon_{0}}{\Delta{t}^{2}}A^{n+1}_{xi+\frac{1}{2},j,k}=-\frac{\epsilon_{0}}{\Delta{t}\Delta{x}}\left(\phi^{n+\frac{1}{2}}_{i+1,j,k}-\phi^{n+\frac{1}{2}}_{i,j,k}-\phi^{n-\frac{1}{2}}_{i+1,j,k}+\phi^{n-\frac{1}{2}}_{i,j,k}\right)\nonumber\\
+\frac{A^{n}_{xi+\frac{1}{2},j,k+1}-2A^{n}_{xi+\frac{1}{2},j,k}+A^{n}_{xi+\frac{1}{2},j,k-1}}{\mu_{0}\Delta{z}^{2}}+\frac{A^{n}_{xi+\frac{1}{2},j+1,k}-2A^{n}_{xi+\frac{1}{2},j,k}+A^{n}_{xi+\frac{1}{2},j-1,k}}{\mu_{0}\Delta{y}^{2}}\nonumber\\
-\frac{A^{n}_{zi+1,j,k+\frac{1}{2}}-A^{n}_{zi+1,j,k-\frac{1}{2}}-A^{n}_{zi,j,k+\frac{1}{2}}+A^{n}_{zi,j,k-\frac{1}{2}}}{\mu_{0}\Delta{x}\Delta{z}}\nonumber\\
-\frac{A^{n}_{yi+1,j+\frac{1}{2},k}-A^{n}_{yi+1,j-\frac{1}{2},k}-A^{n}_{yi,j+\frac{1}{2},k}+A^{n}_{yi,j-\frac{1}{2},k}}{\mu_{0}\Delta{x}\Delta{y}}\nonumber\\
+en^{n}_{i+\frac{1}{2},j+\frac{1}{2},k+\frac{1}{2}}v^{n}_{xi+\frac{1}{2},j,k}+\frac{2\epsilon_{0}}{\Delta{t}^{2}}A^{n}_{xi+\frac{1}{2},j,k}-\frac{\epsilon_{0}}{\Delta{t}^{2}}A^{n-1}_{xi+\frac{1}{2},j,k},\label{eq:42}
\end{eqnarray}
\begin{eqnarray}
\frac{\partial{S_{d}}}{\partial{A^{n}_{yi,j+\frac{1}{2},k}}}=\frac{\partial(L^{n-1{\sim}n}_{d}+L^{n{\sim}n+1}_{d})}{\partial{A^{n}_{yi,j+\frac{1}{2},k}}}\Delta{t}=0\Rightarrow\nonumber\\
\frac{\epsilon_{0}}{\Delta{t}^{2}}A^{n+1}_{yi,j+\frac{1}{2},k}=-\frac{\epsilon_{0}}{\Delta{t}\Delta{y}}\left(\phi^{n+\frac{1}{2}}_{i,j+1,k}-\phi^{n+\frac{1}{2}}_{i,j,k}-\phi^{n-\frac{1}{2}}_{i,j+1,k}+\phi^{n-\frac{1}{2}}_{i,j,k}\right)\nonumber\\
+\frac{A^{n}_{yi+1,j+\frac{1}{2},k}-2A^{n}_{yi,j+\frac{1}{2},k}+A^{n}_{yi-1,j+\frac{1}{2},k}}{\mu_{0}\Delta{x}^{2}}+\frac{A^{n}_{yi,j+\frac{1}{2},k+1}-2A^{n}_{yi,j+\frac{1}{2},k}+A^{n}_{yi,j+\frac{1}{2},k-1}}{\mu_{0}\Delta{z}^{2}}\nonumber\\
-\frac{A^{n}_{xi+\frac{1}{2},j+1,k}-A^{n}_{xi-\frac{1}{2},j+1,k}-A^{n}_{xi+\frac{1}{2},j,k}+A^{n}_{xi-\frac{1}{2},j,k}}{\mu_{0}\Delta{y}\Delta{x}}\nonumber\\
-\frac{A^{n}_{zi,j+1,k+\frac{1}{2}}-A^{n}_{zi,j+1,k-\frac{1}{2}}-A^{n}_{zi,j,k+\frac{1}{2}}+A^{n}_{zi,j,k-\frac{1}{2}}}{\mu_{0}\Delta{y}\Delta{z}}\nonumber\\
+en^{n}_{i+\frac{1}{2},j+\frac{1}{2},k+\frac{1}{2}}v^{n}_{yi,j+\frac{1}{2},k}+\frac{2\epsilon_{0}}{\Delta{t}^{2}}A^{n}_{yi,j+\frac{1}{2},k}-\frac{\epsilon_{0}}{\Delta{t}^{2}}A^{n-1}_{yi,j+\frac{1}{2},k},\label{eq:43}
\end{eqnarray}
\begin{eqnarray}
\frac{\partial{S_{d}}}{\partial{A^{n}_{zi,j,k+\frac{1}{2}}}}=\frac{\partial(L^{n-1{\sim}n}_{d}+L^{n{\sim}n+1}_{d})}{\partial{A^{n}_{zi,j,k+\frac{1}{2}}}}=0\Rightarrow\nonumber\\
\frac{\epsilon_{0}}{\Delta{t}^{2}}A^{n+1}_{zi,j,k+\frac{1}{2}}=-\frac{\epsilon_{0}}{\Delta{t}\Delta{z}}\left(\phi^{n+\frac{1}{2}}_{i,j,k+1}-\phi^{n+\frac{1}{2}}_{i,j,k}-\phi^{n-\frac{1}{2}}_{i,j,k+1}+\phi^{n-\frac{1}{2}}_{i,j,k}\right)\nonumber\\
+\frac{A^{n}_{zi,j+1,k+\frac{1}{2}}-2A^{n}_{zi,j,k+\frac{1}{2}}+A^{n}_{zi,j-1,k+\frac{1}{2}}}{\mu_{0}\Delta{y}^{2}}+\frac{A^{n}_{zi+1,j,k+\frac{1}{2}}-2A^{n}_{zi,j,k+\frac{1}{2}}+A^{n}_{zi-1,j,k+\frac{1}{2}}}{\mu_{0}\Delta{x}^{2}}\nonumber\\
-\frac{A^{n}_{yi,j+\frac{1}{2},k+1}-A^{n}_{yi,j-\frac{1}{2},k+1}-A^{n}_{yi,j+\frac{1}{2},k}+A^{n}_{yi,j-\frac{1}{2},k}}{\mu_{0}\Delta{z}\Delta{y}}\nonumber\\
-\frac{A^{n}_{xi+\frac{1}{2},j,k+1}-A^{n}_{xi-\frac{1}{2},j,k+1}-A^{n}_{xi+\frac{1}{2},j,k}+A^{n}_{xi-\frac{1}{2},j,k}}{\mu_{0}\Delta{z}\Delta{x}}\nonumber\\
+en^{n}_{i+\frac{1}{2},j+\frac{1}{2},k+\frac{1}{2}}v^{n}_{zi,j,k+\frac{1}{2}}+\frac{2\epsilon_{0}}{\Delta{t}^{2}}A^{n}_{zi,j,k+\frac{1}{2}}-\frac{\epsilon_{0}}{\Delta{t}^{2}}A^{n-1}_{zi,j,k+\frac{1}{2}},\label{eq:44}
\end{eqnarray}
\begin{eqnarray}
\frac{\partial{S_{d}}}{\partial\phi^{n+\frac{1}{2}}_{i,j,k}}=\frac{\partial{L^{n{\sim}n+1}_{d}}}{\partial\phi^{n+\frac{1}{2}}_{i,j,k}}\Delta{t}=0\Rightarrow\nonumber\\
\frac{\phi^{n+\frac{1}{2}}_{i+1,j,k}-2\phi^{n+\frac{1}{2}}_{i,j,k}+\phi^{n+\frac{1}{2}}_{i-1,j,k}}{\Delta{x}^{2}}+\frac{\phi^{n+\frac{1}{2}}_{i,j+1,k}-2\phi^{n+\frac{1}{2}}_{i,j,k}+\phi^{n+\frac{1}{2}}_{i,j-1,k}}{\Delta{y}^{2}}+\frac{\phi^{n+\frac{1}{2}}_{i,j,k+1}-2\phi^{n+\frac{1}{2}}_{i,j,k}+\phi^{n+\frac{1}{2}}_{i,j,k-1}}{\Delta{z}^{2}}\nonumber\\
+\frac{A^{n+1}_{xi+\frac{1}{2},j,k}-A^{n}_{xi+\frac{1}{2},j,k}-A^{n+1}_{xi-\frac{1}{2},j,k}+A^{n}_{xi-\frac{1}{2},j,k}}{\Delta{t}\Delta{x}}\nonumber\\
+\frac{A^{n+1}_{yi,j+\frac{1}{2},k}-A^{n}_{yi,j+\frac{1}{2},k}-A^{n+1}_{yi,j-\frac{1}{2},k}+A^{n}_{yi,j-\frac{1}{2},k}}{\Delta{t}\Delta{y}}\nonumber\\
+\frac{A^{n+1}_{zi,j,k+\frac{1}{2}}-A^{n}_{zi,j,k+\frac{1}{2}}-A^{n+1}_{zi,j,k-\frac{1}{2}}+A^{n}_{zi,j,k-\frac{1}{2}}}{\Delta{t}\Delta{z}}=-\frac{e}{\epsilon_{0}}\left(n^{n}_{i+\frac{1}{2},j+\frac{1}{2},k+\frac{1}{2}}-n_{0}\right),\label{eq:45}
\end{eqnarray}
\begin{eqnarray}
\frac{\partial{S_{d}}}{\partial\alpha^{n+\frac{1}{2}}_{i,j,k}}=\frac{\partial{L^{n{\sim}n+1}_{d}}}{\partial\alpha^{n+\frac{1}{2}}_{i,j,k}}\Delta{t}=0\Rightarrow\nonumber\\
n^{n+1}_{i+\frac{1}{2},j+\frac{1}{2},k+\frac{1}{2}}+\frac{\Delta{t}}{\Delta{x}}\left(n^{n+1}_{i+\frac{1}{2},j+\frac{1}{2},k+\frac{1}{2}}v^{n+1}_{xi+\frac{1}{2},j,k}-n^{n+1}_{i-\frac{1}{2},j+\frac{1}{2},k+\frac{1}{2}}v^{n+1}_{xi-\frac{1}{2},j,k}\right)\nonumber\\
+\frac{\Delta{t}}{\Delta{y}}\left(n^{n+1}_{i+\frac{1}{2},j+\frac{1}{2},k+\frac{1}{2}}v^{n+1}_{yi,j+\frac{1}{2},k}-n^{n+1}_{i+\frac{1}{2},j-\frac{1}{2},k+\frac{1}{2}}v^{n+1}_{yi,j-\frac{1}{2},k}\right)\nonumber\\
+\frac{\Delta{t}}{\Delta{z}}\left(n^{n+1}_{i+\frac{1}{2},j+\frac{1}{2},k+\frac{1}{2}}v^{n+1}_{zi,j,k+\frac{1}{2}}-n^{n+1}_{i+\frac{1}{2},j+\frac{1}{2},k-\frac{1}{2}}v^{n+1}_{zi,j,k-\frac{1}{2}}\right)=n^{n}_{i+\frac{1}{2},j+\frac{1}{2},k+\frac{1}{2}},\label{eq:46}
\end{eqnarray}
\begin{eqnarray}
\frac{\partial{S_{d}}}{\partial\lambda^{n}_{i+\frac{1}{2},j+\frac{1}{2},k+\frac{1}{2}}}=\frac{\partial{L^{n{\sim}n+1}_{d}}}{\partial\lambda^{n}_{i+\frac{1}{2},j+\frac{1}{2},k+\frac{1}{2}}}\Delta{t}=0\Rightarrow\nonumber\\
\mu^{n+\frac{1}{2}}_{i,j,k}=\mu^{n-\frac{1}{2}}_{i,j,k}-\frac{\Delta{t}}{\Delta{x}}v^{n}_{xi+\frac{1}{2},j,k}\left(\mu^{n-\frac{1}{2}}_{i+1,j,k}-\mu^{n-\frac{1}{2}}_{i,j,k}\right)\nonumber\\
-\frac{\Delta{t}}{\Delta{y}}v^{n}_{yi,j+\frac{1}{2},k}\left(\mu^{n-\frac{1}{2}}_{i,j+1,k}-\mu^{n-\frac{1}{2}}_{i,j,k}\right)-\frac{\Delta{t}}{\Delta{z}}v^{n}_{zi,j,k+\frac{1}{2}}\left(\mu^{n-\frac{1}{2}}_{i,j,k+1}-\mu^{n-\frac{1}{2}}_{i,j,k}\right),\label{eq:47}
\end{eqnarray}
\begin{eqnarray}
\frac{\partial{S_{d}}}{\partial\mu^{n-\frac{1}{2}}_{i,j,k}}=\frac{\partial(L^{n-1{\sim}n}_{d}+L^{n{\sim}n+1}_{d})}{\partial\mu^{n-\frac{1}{2}}_{i,j,k}}\Delta{t}=0\Rightarrow\nonumber\\
\lambda^{n}_{i+\frac{1}{2},j+\frac{1}{2},k+\frac{1}{2}}+\frac{\Delta{t}}{\Delta{x}}\left(\lambda^{n}_{i+\frac{1}{2},j+\frac{1}{2},k+\frac{1}{2}}v^{n}_{xi+\frac{1}{2},j,k}-\lambda^{n}_{i-\frac{1}{2},j+\frac{1}{2},k+\frac{1}{2}}v^{n}_{xi-\frac{1}{2},j,k}\right)\nonumber\\
+\frac{\Delta{t}}{\Delta{y}}\left(\lambda^{n}_{i+\frac{1}{2},j+\frac{1}{2},k+\frac{1}{2}}v^{n}_{yi,j+\frac{1}{2},k}-\lambda^{n}_{i+\frac{1}{2},j-\frac{1}{2},k+\frac{1}{2}}v^{n}_{yi,j-\frac{1}{2},k}\right)\nonumber\\
+\frac{\Delta{t}}{\Delta{z}}\left(\lambda^{n}_{i+\frac{1}{2},j+\frac{1}{2},k+\frac{1}{2}}v^{n}_{zi,j,k+\frac{1}{2}}-\lambda^{n}_{i+\frac{1}{2},j+\frac{1}{2},k-\frac{1}{2}}v^{n}_{zi,j,k-\frac{1}{2}}\right)=\lambda^{n-1}_{i+\frac{1}{2},j+\frac{1}{2},k+\frac{1}{2}}.\label{eq:48}
\end{eqnarray}

\begin{figure}[htbp]
\centerline{\includegraphics[width=8cm,height=5.6cm]{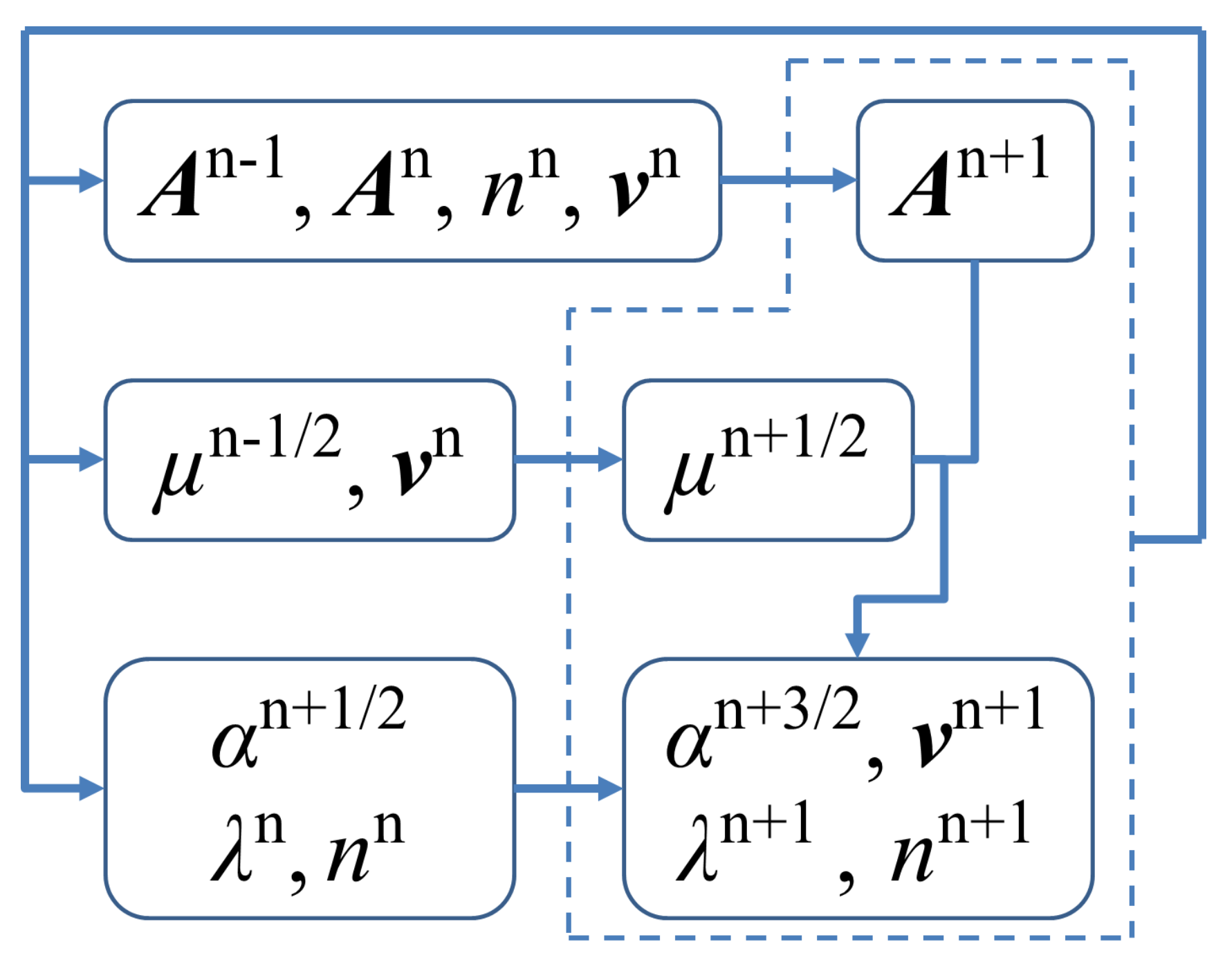}}
\caption{The complete iteration for the variational schemes of the hydrodynamic-electrodynamic model.}
\label{fig:3}
\end{figure}

Eqs.~\eqref{eq:39}-\eqref{eq:41} and \eqref{eq:45} are constraints that restrict the solution manifold. Both 
initialization and evolution should be restricted by these numerical constraints. Eqs.~\eqref{eq:42}-\eqref{eq:44} are 
explicit schemes used for updating the gauge field components. We emphasize that in a DEC framework and a rectangular 
lattice, the variational schemes \eqref{eq:42}-\eqref{eq:44} of Maxwell's equations are equal to the traditional FDTD 
method. Based on DEC, the electromagnetic fields in a rectangular lattice are defined as,
\begin{eqnarray}
E^{n+\frac{1}{2}}_{xi+\frac{1}{2},j,k}=-\frac{A^{n+1}_{xi+\frac{1}{2},j,k}-A^{n}_{xi+\frac{1}{2},j,k}}{\Delta{t}},\label{eq:49}
\end{eqnarray}
\begin{eqnarray}
E^{n+\frac{1}{2}}_{yi,j+\frac{1}{2},k}=-\frac{A^{n+1}_{yi,j+\frac{1}{2},k}-A^{n}_{yi,j+\frac{1}{2},k}}{\Delta{t}},\label{eq:50}
\end{eqnarray}
\begin{eqnarray}
E^{n+\frac{1}{2}}_{zi,j,k+\frac{1}{2}}=-\frac{A^{n+1}_{zi,j,k+\frac{1}{2}}-A^{n}_{zi,j,k+\frac{1}{2}}}{\Delta{t}},\label{eq:51}
\end{eqnarray}
\begin{eqnarray}
\mu_{0}H^{n}_{xi,j+\frac{1}{2},k+\frac{1}{2}}=\frac{A^{n}_{zi,j+1,k+\frac{1}{2}}-A^{n}_{zi,j,k+\frac{1}{2}}}{\Delta{y}}-\frac{A^{n}_{yi,j+\frac{1}{2},k+1}-A^{n}_{yi,j+\frac{1}{2},k}}{\Delta{z}},\label{eq:52}
\end{eqnarray}
\begin{eqnarray}
\mu_{0}H^{n}_{yi+\frac{1}{2},j,k+\frac{1}{2}}=\frac{A^{n}_{xi+\frac{1}{2},j,k+1}-A^{n}_{xi+\frac{1}{2},j,k}}{\Delta{z}}-\frac{A^{n}_{zi+1,j,k+\frac{1}{2}}-A^{n}_{zi,j,k+\frac{1}{2}}}{\Delta{x}},\label{eq:53}
\end{eqnarray}
\begin{eqnarray}
\mu_{0}H^{n}_{zi+\frac{1}{2},j+\frac{1}{2},k}=\frac{A^{n}_{yi+1,j+\frac{1}{2},k}-A^{n}_{yi,j+\frac{1}{2},k}}{\Delta{x}}-\frac{A^{n}_{xi+\frac{1}{2},j+1,k}-A^{n}_{xi+\frac{1}{2},j,k}}{\Delta{y}},\label{eq:54}
\end{eqnarray}
where the temporal gauge $\phi^{n+\frac{1}{2}}_{i,j,k}=0$ has been adopted explicitly. Then substituting Eqs.~\eqref{eq:49}-\eqref{eq:54} 
into Eqs.~\eqref{eq:42}-\eqref{eq:44}, we obtain,
\begin{eqnarray}
E^{n+\frac{1}{2}}_{xi+\frac{1}{2},j,k}&=&E^{n-\frac{1}{2}}_{xi+\frac{1}{2},j,k}-\frac{\Delta{t}}{\epsilon_{0}}J^{n}_{xi+\frac{1}{2},j,k}+\frac{\Delta{t}}{\epsilon_{0}\Delta{y}}\left(H^{n}_{zi+\frac{1}{2},j+\frac{1}{2},k}-H^{n}_{zi+\frac{1}{2},j-\frac{1}{2},k}\right)\nonumber\\
& &-\frac{\Delta{t}}{\epsilon_{0}\Delta{z}}\left(H^{n}_{yi+\frac{1}{2},j,k+\frac{1}{2}}-H^{n}_{yi+\frac{1}{2},j,k-\frac{1}{2}}\right),\label{eq:55}
\end{eqnarray}
\begin{eqnarray}
E^{n+\frac{1}{2}}_{yi,j+\frac{1}{2},k}&=&E^{n-\frac{1}{2}}_{yi,j+\frac{1}{2},k}-\frac{\Delta{t}}{\epsilon_{0}}J^{n}_{yi,j+\frac{1}{2},k}+\frac{\Delta{t}}{\epsilon_{0}\Delta{z}}\left(H^{n}_{xi,j+\frac{1}{2},k+\frac{1}{2}}-H^{n}_{xi,j+\frac{1}{2},k-\frac{1}{2}}\right)\nonumber\\
& &-\frac{\Delta{t}}{\epsilon_{0}\Delta{x}}\left(H^{n}_{zi+\frac{1}{2},j+\frac{1}{2},k}-H^{n}_{zi-\frac{1}{2},j+\frac{1}{2},k}\right),\label{eq:56}
\end{eqnarray}
\begin{eqnarray}
E^{n+\frac{1}{2}}_{zi,j,k+\frac{1}{2}}&=&E^{n-\frac{1}{2}}_{zi,j,k+\frac{1}{2}}-\frac{\Delta{t}}{\epsilon_{0}}J^{n}_{zi,j,k+\frac{1}{2}}+\frac{\Delta{t}}{\epsilon_{0}\Delta{x}}\left(H^{n}_{yi+\frac{1}{2},j,k+\frac{1}{2}}-H^{n}_{yi-\frac{1}{2},j,k+\frac{1}{2}}\right)\nonumber\\
& &-\frac{\Delta{t}}{\epsilon_{0}\Delta{y}}\left(H^{n}_{xi,j+\frac{1}{2},k+\frac{1}{2}}-H^{n}_{xi,j-\frac{1}{2},k+\frac{1}{2}}\right),\label{eq:57}
\end{eqnarray}
where $(J_{x},J_{y},J_{z})=(env_{x},env_{y},env_{z})$ is the current density. Eqs.~\eqref{eq:55}-\eqref{eq:57} have the 
usual form of the standard FDTD method constructed in a Yee lattice \cite{Yee}. It is well known that the standard FDTD 
is symplectic, which means the schemes \eqref{eq:42}-\eqref{eq:44} of Maxwell's equations are also symplectic. We will 
give a brief proof of this corollary in the following. Eq.~\eqref{eq:47} is an explicit scheme used for updating Lin's 
constraint field. Eqs.~\eqref{eq:38}, \eqref{eq:46}, and \eqref{eq:48} are implicit schemes of the electron density and 
Lagrangian multipliers. It can be seen that Eqs.~\eqref{eq:38}-\eqref{eq:41}, \eqref{eq:46}, and \eqref{eq:48} make up 
an implicit cubic nonlinear algebraic system that can be solved for updating the electron density, velocity components, 
and multipliers. Effective and efficient nonlinear iterative methods and linear solvers are needed to solve the nonlinear 
algebraic equations. The complete iteration for the variational schemes is shown in Fig.\ref{fig:3}.

A very good property of the variational scheme is the conservation of Lagrangian structure generated by the discrete 
Lagrangian \eqref{eq:37} $L^{n{\sim}n+1}_{d}(q^{n},q^{n+1})$, where 
$q^{n}=(n^{n}_{J},\bm{v}^{n}_{J},\bm{A}^{n}_{J},\phi^{n+\frac{1}{2}}_{J},\alpha^{n+\frac{1}{2}}_{J},\lambda^{n}_{J},\mu^{n-\frac{1}{2}}_{J})$ and the 
subscript $J$ traverses all lattice points. We define two 1-forms,
\begin{eqnarray}
\bm{\theta}^{+n{\sim}n+1}=\frac{\partial}{\partial{q^{n+1}_{i}}}L^{n{\sim}n+1}_{d}\left(q^{n},q^{n+1}\right)\cdot\bm{{\rm{d}}}q^{n+1}_{i},\label{eq:58}
\end{eqnarray}
\begin{eqnarray}
\bm{\theta}^{-n{\sim}n+1}=-\frac{\partial}{\partial{q^{n}_{i}}}L^{n{\sim}n+1}_{d}\left(q^{n},q^{n+1}\right)\cdot\bm{{\rm{d}}}q^{n}_{i},\label{eq:59}
\end{eqnarray}
which form a partition of the exterior derivative of the discrete Lagrangian 
$\bm{{\rm{d}}}L^{n{\sim}n+1}_{d}=\bm{\theta}^{+n{\sim}n+1}-\bm{\theta}^{-n{\sim}n+1}$. Then the Lagrangian noncanonical 
structure can be given as a closed 2-form 
$\bm{\Omega}^{n{\sim}n+1}_{d}=\bm{{\rm{d}}}\bm{\theta}^{+n{\sim}n+1}=\bm{{\rm{d}}}\bm{\theta}^{-n{\sim}n+1}$ as,
\begin{eqnarray}
\bm{\Omega}^{n{\sim}n+1}_{d}=\frac{\partial^{2}}{\partial{q^{n}_{i}}\partial{q^{n+1}_{j}}}L^{n{\sim}n+1}_{d}\left(q^{n},q^{n+1}\right)\cdot\bm{{\rm{d}}}q^{n}_{i}\wedge\bm{{\rm{d}}}q^{n+1}_{j}.\label{eq:60}
\end{eqnarray}
This discrete geometric structure is symplectic. By using the discrete dynamical equations \eqref{eq:38}-\eqref{eq:48} 
and taking the exterior derivative of action $S_{d}$, we obtain,
\begin{eqnarray}
\bm{{\rm{d}}}S_{d}=\bm{\theta}^{+N{\sim}N+1}-\bm{\theta}^{-0{\sim}1}.\label{eq:61}
\end{eqnarray}
Eq.~\eqref{eq:61} means that one more exterior derivative of this 1-form leads to,
\begin{eqnarray}
\bm{\Omega}^{N{\sim}N+1}_{d}=\bm{\Omega}^{0{\sim}1}_{d}.\label{eq:62}
\end{eqnarray}
The hydrodynamic-Maxwell system generated by Eq.~\eqref{eq:10} is naturally equipped with a continuous Lagrangian 
symplectic structure. The conservation of discrete Lagrangian symplectic structure indicates the discrete variational 
principle generates a self-consistent finite-dimensional dynamical system that is a good analog of the continuous 
system. We should emphasize that preserving the discrete symplectic structure in the extended system (with Lagrangian 
multipliers) is not a necessary and sufficient condition for preserving the geometric structures in the original system. 
When the dynamical system is completely integrable, the Kolmogorov-Arnold-Moser (KAM) theorem states that a weak 
perturbation in the Hamiltonian will not break the invariant tori of the system, which is a theoretical root to ensure 
the good numerical properties of the structure-preserving algorithms \cite{Hairer1}. When it comes to a general dynamical 
system, although the KAM theorem is not always available, many numerical experiments show that the structure-preserving 
algorithms still exhibit good behaviors in long-term simulations \cite{Feng2,Hairer1}.

It can be directly examined that the gauge and translation symmetries are also preserved in the discrete dynamical 
equations. By introducing an arbitrary 0-form $\psi$, we can define the discrete gauge transformation, 
\begin{eqnarray}
\phi^{'n+\frac{1}{2}}_{i,j,k}=\phi^{n+\frac{1}{2}}_{i,j,k}-\frac{\psi^{n+1}_{i,j,k}-\psi^{n}_{i,j,k}}{\Delta{t}},\label{eq:gt1}\\
A^{'n}_{xi+\frac{1}{2},j,k}=A^{n}_{xi+\frac{1}{2},j,k}+\frac{\psi^{n}_{i+1,j,k}-\psi^{n}_{i,j,k}}{\Delta{x}},\label{eq:gt2}\\
A^{'n}_{yi,j+\frac{1}{2},k}=A^{n}_{yi,j+\frac{1}{2},k}+\frac{\psi^{n}_{i,j+1,k}-\psi^{n}_{i,j,k}}{\Delta{y}},\label{eq:gt3}\\
A^{'n}_{zi,j,k+\frac{1}{2}}=A^{n}_{zi,j,k+\frac{1}{2}}+\frac{\psi^{n}_{i,j,k+1}-\psi^{n}_{i,j,k}}{\Delta{z}}.\label{eq:gt4}
\end{eqnarray}
Substituting Eqs.~\eqref{eq:gt1}-\eqref{eq:gt4} into the discrete Lagrangian density \eqref{eq:35}, and summing over 
it on a universal discrete space-time manifold, we can obtain the gauge-invariant discrete action. Based on the discrete 
Noether's theorem, it gives the discrete charge conservation law \cite{Marsden}. Additionally, the discrete Lagrangian 
density \eqref{eq:35} is space-time-coordinate-independent, which means that the discrete action is translation-invariant. 
Based on the discrete Noether's theorem, the Lagrangian momentum maps preserve the general energy-momentum density \cite{Marsden}. 
The advantages of variational schemes in long-term simulations are ensured by the conservation of the discrete 
symplectic 2-form, gauge, and translation symmetries, which lead to the conserved quantities having long-term conservation 
and accuracy in simulations.

The variational schemes constructed here are recognized as a structure-preserving Eulerian algorithm, which is 
convenient to implement and parallelize. There are also other types of structure-preserving algorithms constructed for 
hydrodynamic-electrodynamic systems, such as the symplectic Smoothed-Particle-Hydrodynamics (SPH) method used for 
simulating the double-fluid model of plasmas \cite{JXiao}. The symplectic SPH method is a Hybrid-Eulerian-Lagrangian 
(HEL) algorithm which avoids constraints and is suitable for simulating plasma waves. But the metaparticle interpolation 
is tedious and time-consuming. The Eulerian form of the variational schemes in this work without tedious interpolation is 
more suitable for nano-optics, although the constraints may have a singular performance under a few conditions. As the 
Lagrangian multipliers are pure mathematical variables, their initialization is only determined by Eqs.~\eqref{eq:39}
-\eqref{eq:41}. After the physical variables are initialized self-consistently, we can obtain the initial conditions of 
multipliers and Lin's constraint field by solving Eqs.~\eqref{eq:39}-\eqref{eq:41}.

\subsection{Algebraic solvers\label{sec:3-3}}
To implement the variational scheme, the solving procedure for nonlinear algebraic equations \eqref{eq:38}-\eqref{eq:41}, 
\eqref{eq:46}, and \eqref{eq:48} is a core technique. In this work, we introduce the Newton-Krylov-type methods as 
primary algebraic solvers \cite{Brown}. As the shell of a Newton-Krylov-type method, the Newton-Raphson algorithm is 
used as the basic nonlinear iterative method which approximates and corrects the algebraic system in every numerical step.

The nonlinear equations \eqref{eq:38}-\eqref{eq:41}, \eqref{eq:46}, and \eqref{eq:48} can rewritten in matrix form as,
\begin{eqnarray}
\bm{F}\left(\bm{X}^{n+1}\right)=0,\label{eq:63}
\end{eqnarray}
where $\bm{F}$ is a well-defined nonlinear vector function, 
$\bm{X}=(\alpha_{J},v_{xJ},v_{yJ},v_{zJ},n_{J},\lambda_{J})^{T}$, and the subscript $J$ traverses all 
lattice points. Then the Newton-Raphson iteration is given as,
\begin{eqnarray}
\bm{X}^{*}=\bm{X}-\bm{J}^{-1}_{F}\left(\bm{X}\right)\bm{F}\left(\bm{X}\right),\label{eq:64}
\end{eqnarray}
where $\bm{X}^{*}$ indicates new variables, and the Jacobian $\bm{J}_{F}$ should be updated in every iteration step. 
The detailed Jacobian elements can be found in App.~\ref{sec:app}. The criterion $||\bm{X}^{*}-\bm{X}||/||\bm{X}||<\varepsilon$ 
is used to cutoff the iteration, where $\varepsilon$ is a specified sufficiently small value.

During every iteration step, we face a Jacobian inversion, which means a linear algebraic matrix equation needs to be 
solved. Based on the Krylov subspace theory, there are many efficient linear solvers. For example, the Generalized-Minimum-Residual 
(GMRES) method, the Incomplete-Cholesky-Conjugate-Gradient (ICCG) method, and the BICGSTAB method, have been constructed 
to solve a large sparse matrix equation \cite{Saad,Meijerink,Vorst}. In this work, the BICGSTAB method is introduced as 
the basic linear solver, because of its efficiency, stability, and parallel ability \cite{Vorst}. An alternative approach 
to solve the linear equations generated in nonlinear iteration is the Jacobian-Free Newton-Krylov (JFNK) method, which 
replaces $\bm{J}_{F}\bm{X}^{*}$ with $\bm{J}_{F}\bm{X}^{*}\approx[\bm{J}_{F}(\bm{X}+\xi\bm{X}^{*})-\bm{J}_{F}(\bm{X})]/\xi$, 
where $\xi$ is a small perturbation \cite{Brown,Knoll}. A good feature of the JFNK method is that the Jacobian-vector 
product can be probed approximately without forming and storing the Jacobian elements. In this work, we just use the 
Newton-BICGSTAB iteration method to solve the nonlinear system, as the Jacobian can be exactly derived conveniently 
(App.~\ref{sec:app}). The JFNK approximation will be taken into consideration to accelerate the simulation in future work.

\section{Numerical Experiments\label{sec:4}}

\subsection{Bulk Plasmon\label{sec:4-1}}
\begin{figure}[htbp]
\centerline{\includegraphics[width=9.2cm,height=6.4cm]{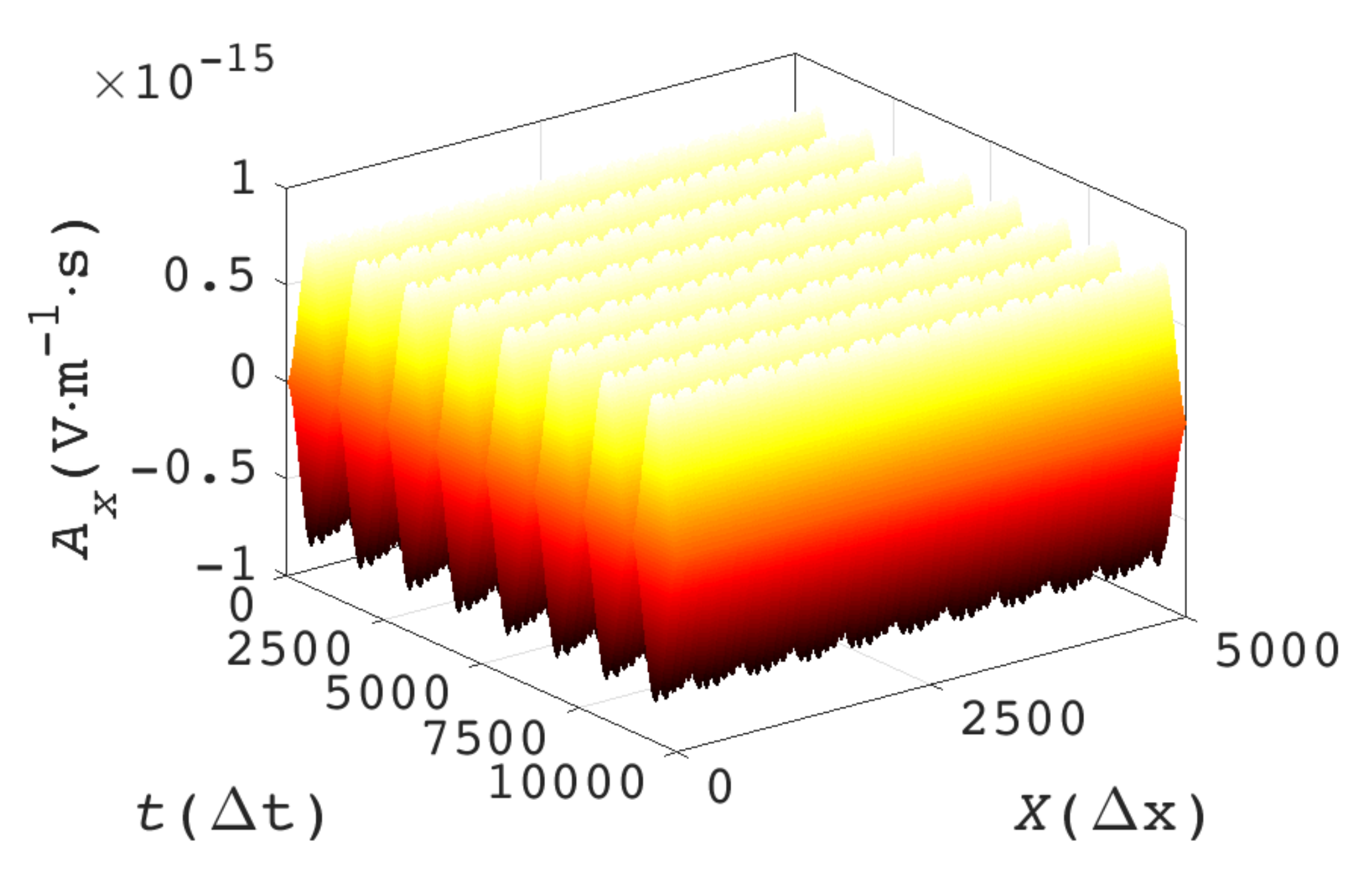}}
\caption{The evolution of the gauge field in the simulation of bulk plasmon oscillation.}
\label{fig:4}
\end{figure}

\begin{figure}[htbp]
\centerline{\includegraphics[width=8.7cm,height=6cm]{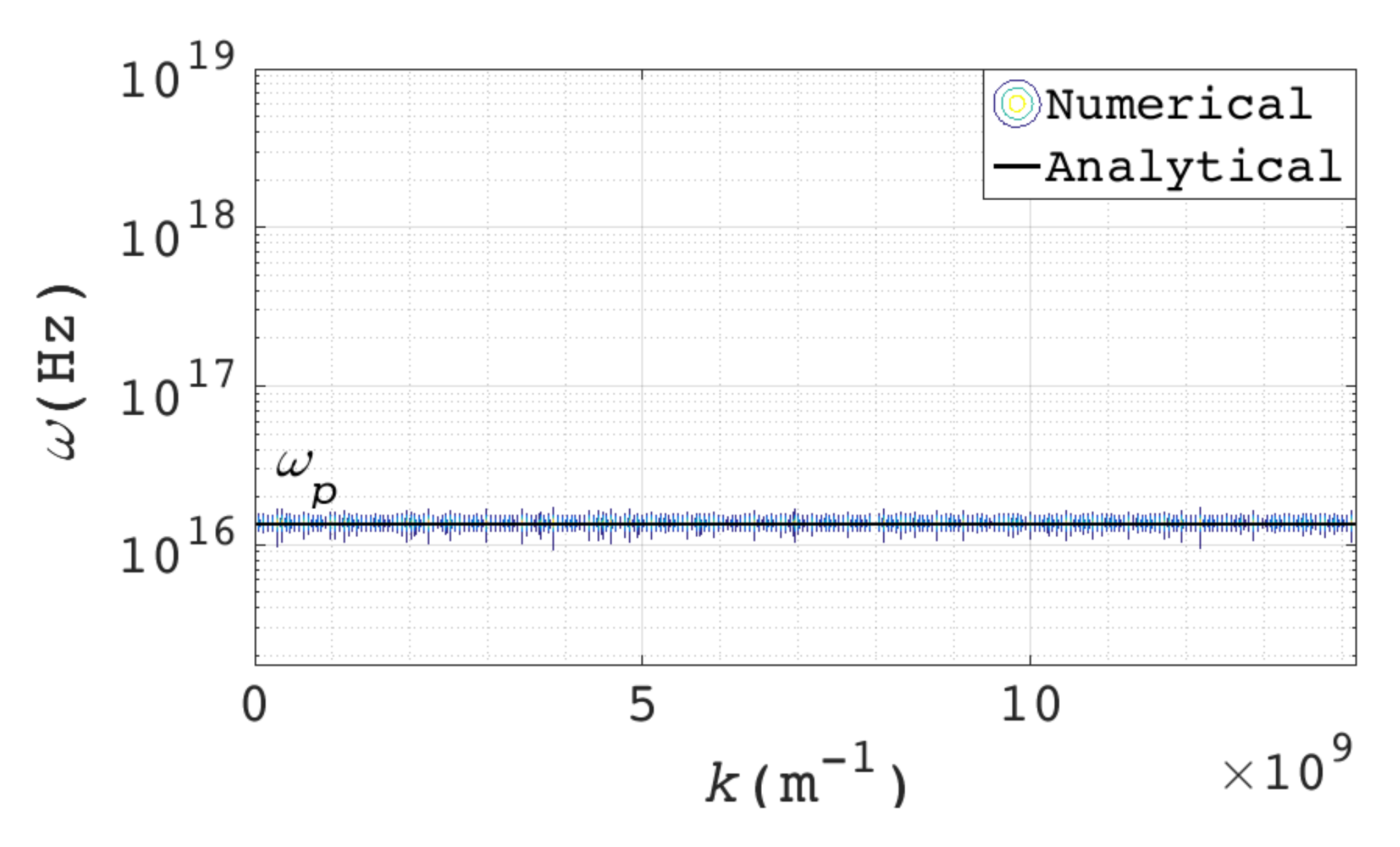}}
\caption{Bulk plasmon dispersion relations: the numerical result (contour plot) versus the analytical one (solid line).}
\label{fig:5}
\end{figure}

The first numerical experiment implemented in this work is the one-dimensional (1-D) bulk plasmon oscillation. This 
simulation is a numerical benchmark that is used to verify the variational code. In this 1-D simulation, the metal is 
specified as silver, which means the the plasma frequency $\omega_{p}=1.37\times10^{16}$ $\mathrm{Hz}$ (for more details, 
see Sec.\ref{sec:4-2}). Then the spatial step is chosen as $0.01c/\omega_{p}$, and the temporal step is determined 
by the Courant-Friedrichs-Lewy (CFL) constraint, where $CFL=0.5$. The numerical simulation domain is a 5000 1-D lattice, 
and the Message Passing Interface (MPI) is used as a parallel strategy. At the initial time, a random perturbation of 
gauge field is introduced as the initial condition. After 10000 steps simulation, the dispersion relation can be 
reconstructed by taking the Fast Fourier Transform (FFT) of the gauge field. Fig.~\ref{fig:4} shows the evolution of 
the gauge field in the simulation, where the plasmon oscillation with frequency $\omega_{p}$ can be recognized obviously. 
Fig.~\ref{fig:5} plots the numerical dispersion relation versus the analytical one. The numerical result (contour plot) 
has a good consistency with the analytical dispersion relation (solid line) in the truncation region $k\in$ [0.00057,1.4] 
$\times10^{10}$m$^{-1}$. It shows that the accurate linear response has been reached over a wide range of the spectrum, 
which means the simulated plasmonic system is physically correct.

\begin{figure}[htbp]
\centerline{\includegraphics[width=9.2cm,height=5.8cm]{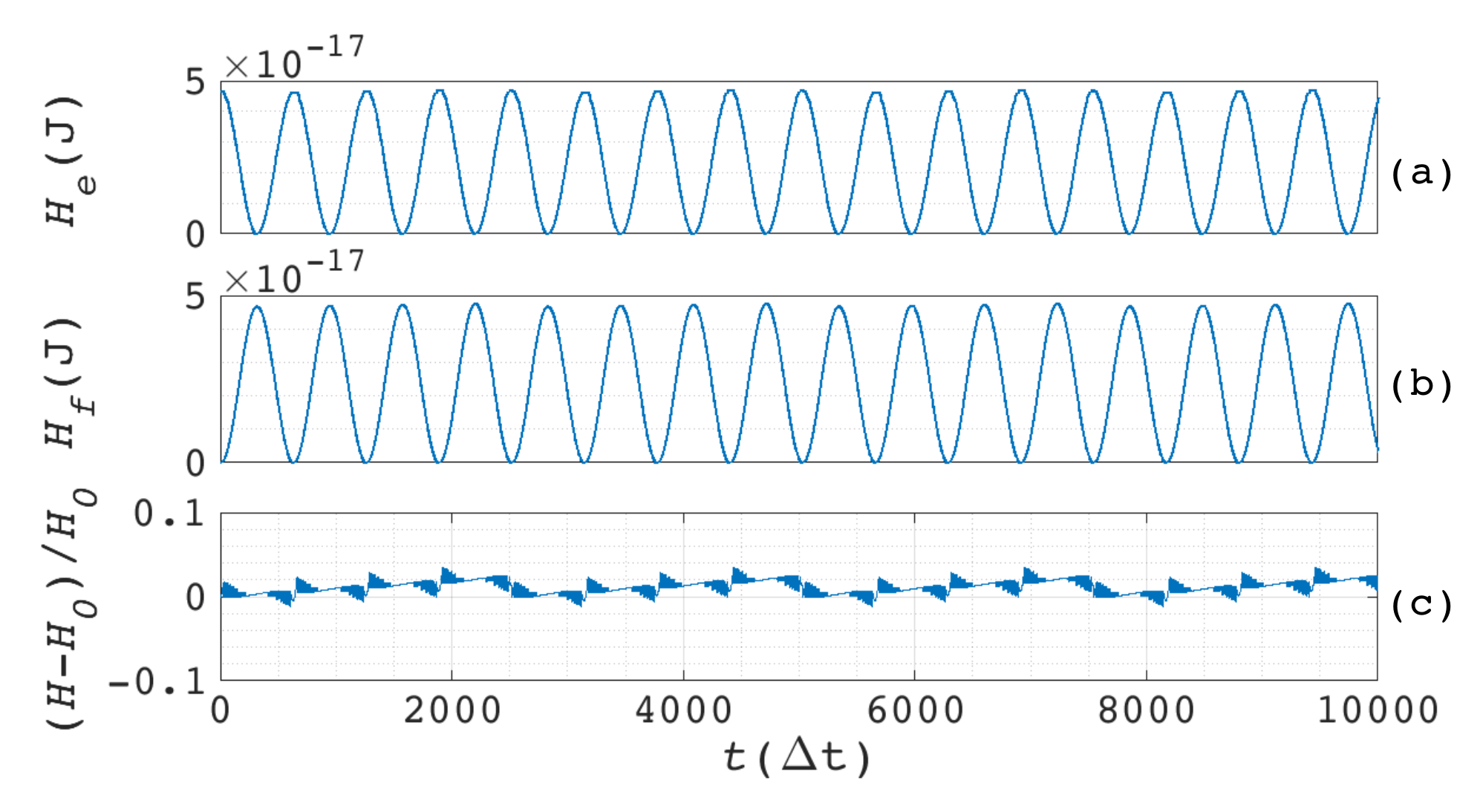}}
\caption{Numerical error of the Hamiltonian in the simulation: (a) the Hamiltonian of losslossless free electron gas; 
(b) the Hamiltonian of self-consistent gauge field; (c) the total Hamiltonian error. The unit of Hamiltonian is Joule.}
\label{fig:6}
\end{figure}

To illustrate the good property of variational schemes in secular simulations, we plot the evolution of numerical error 
of total Hamiltonian, which is shown in Fig.~\ref{fig:6} (c). After a long term simulation, the numerical error is 
bounded by a very small value without coherent accumulation. From the sub-Fig.~\ref{fig:6} (a) and (b), we can find 
that the perturbation energy is exchanged cycle by cycle with frequency $\omega_{p}$ between the electron and field 
components during the oscillation, which drives the plasmonic motion.

The result of first numerical experiment provide us with a basic verification of the numerical code used for simulating 
plasmonic phenomena. The advantage in conservation shown by numerical error is a footstone of secular simulation for a 
nonlinear system.

\subsection{SPPs\label{sec:4-2}}
\begin{figure}[htbp]
\centerline{\includegraphics[width=8cm,height=6cm]{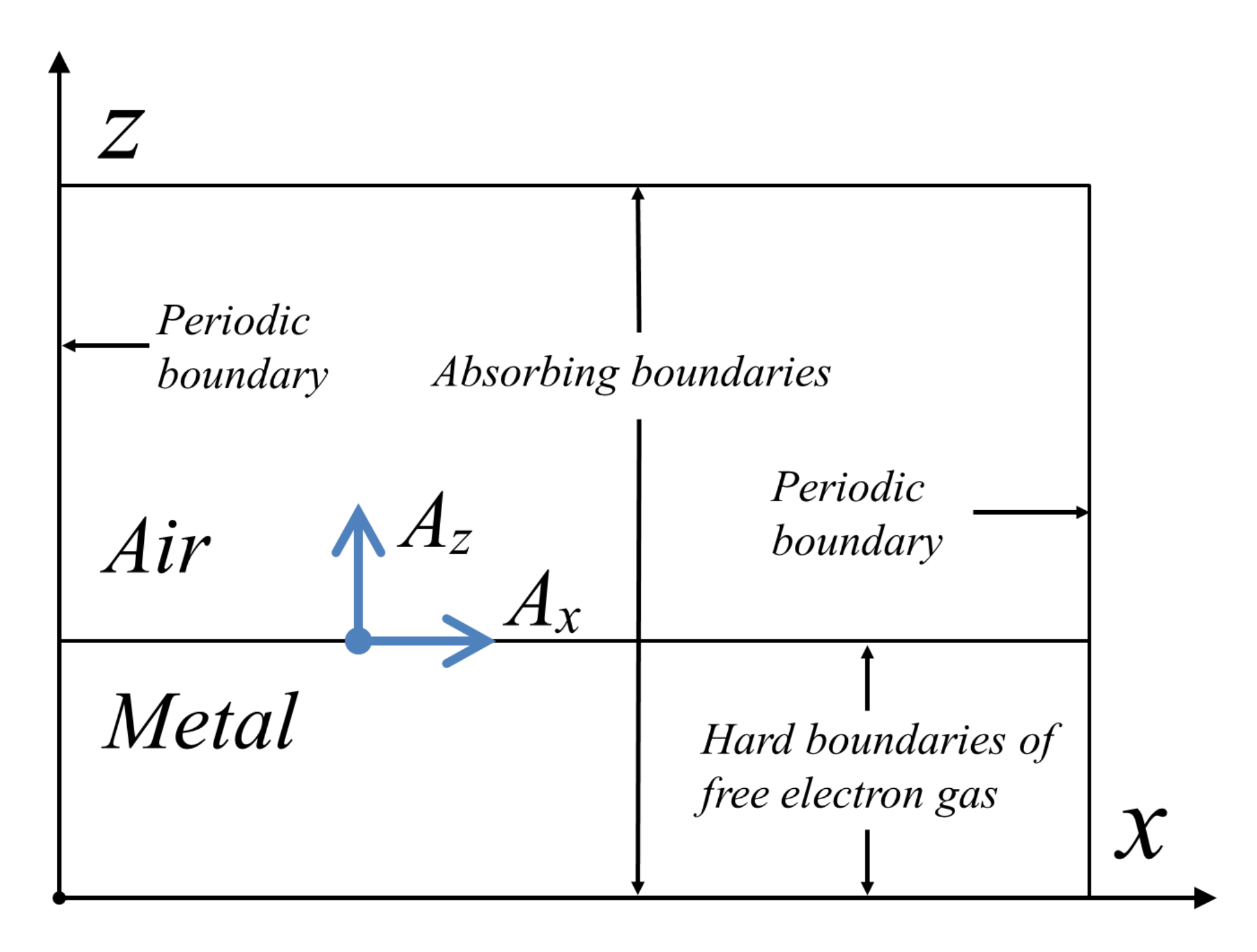}}
\caption{The 2-D SPP configuration used in the simulations.}
\label{fig:7}
\end{figure}

A typical SPPs configuration is the infinite interface between metal and air. When it comes to silver, the average 
electron number density $n_{0}\approx5.90\times10^{28}$ $\mathrm{m}^{-3}$, and then the plasma frequency 
$\omega_{p}=\sqrt{n_{0}e^2/\epsilon_{0}m}=1.37\times10^{16}$ $\mathrm{Hz}$. Based on the Drude model, the metallic 
permittivity can be given as $\epsilon(\omega)=1-\omega^{2}_{p}/\omega^{2}$. The SPPs dispersion relation can be 
analytically given as $k_{x}=k_{0}\sqrt{\epsilon/(\epsilon+1)}$, where $k_{0}$ is the vacuum wave number, and then the 
SPR frequency $\omega_{sp}=0.97\times10^{16}$ $\mathrm{Hz}$. Fig.~\ref{fig:7} shows the 2-D SPP configuration used in
the simulations. The numerical simulation domain is a $200\times150$ (metal $200\times50$) uniform 2-D lattice, where 
the periodic boundaries are used in the x-direction, the absorbing boundaries of gauge field components are used in the 
z-direction, and the hard boundaries of free electron gas are used in the z-direction. The TM mode SPPs with several vacuum 
wave lengthes (300, 280, 260, 240, 220, 200) nm are simulated. In each simulation, the spatial step $\Delta{x}=\Delta{z}$ 
is $0.05\times{2\pi}/k_{x}$, the temporal step $\Delta{t}$ is determined by the CFL constraint $CFL=0.5$, and the total 
simulation time is 4000 steps. To effectively implement the simulations, the MPI is used to parallel the code.

\begin{figure}[htbp]
\centerline{\includegraphics[width=8cm,height=6.7cm]{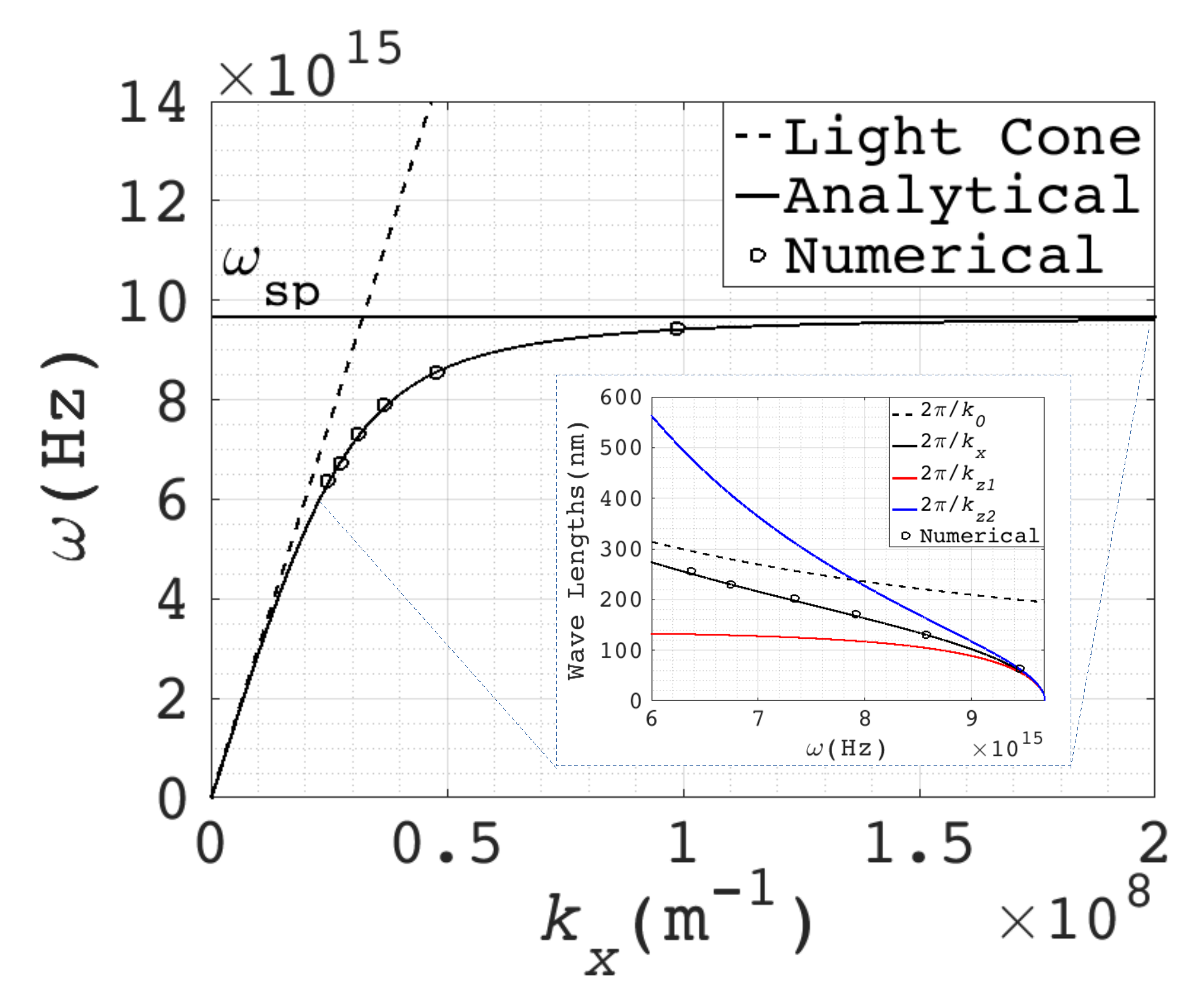}}
\caption{SPPs dispersion relations. Analytical vs numerical. Numerical results obtained at given wave lengths 
have a good consistency with the analytical curve. The inset plots the wave lengths and decay lengths at simulation 
region, which illustrates the sub-wavelength and energy localization of SPPs.}
\label{fig:8}
\end{figure}

\begin{figure}[htbp]
\centerline{\includegraphics[width=9.2cm,height=5.8cm]{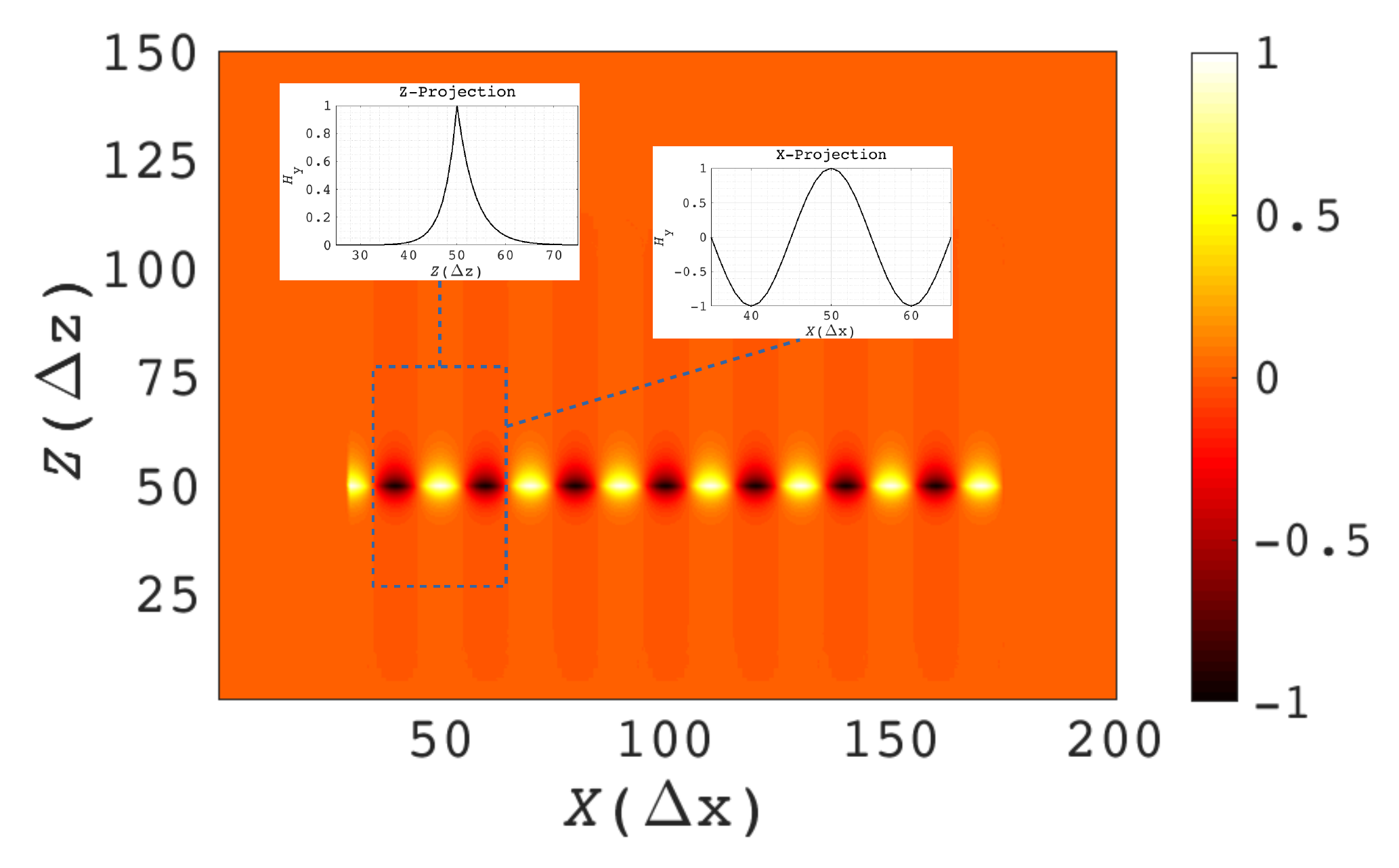}}
\caption{The slice of normalized SPPs field $H_{y}$ obtained by simulations. The vacuum wave length $\lambda_{0}=240$ nm. 
The insets plot the X- and Z- projections of $H_{y}$ at the region [35,65]$\times$[25,75]. It shows that the 
electromagnetic field of SPPs is strongly bounded by the interface, and the mode has a sub-wavelength spatial structure.}
\label{fig:9}
\end{figure}

Fig.~\ref{fig:8} shows the dispersion relation at given wave lengths obtained by simulations. By comparing the 
numerical results (circle marks) with analytical dispersion curve, we find that the simulations can reproduce 
characteristic dispersion relation of SPPs from near light cone to SPR regions. The resonance of wave means super 
resolution which is a core property of near field optics. The inset of Fig.~\ref{fig:8} plots the wave lengths 
and decay lengths in the simulation region. It shows that the SPPs wave length (black line) and metallic decay length 
(red line) are always less than the vacuum wave length (dashed line), but the air decay length (blue line) is less than 
the vacuum wave length only in the near SPR region ($\omega>$7.9$\times10^{15}$Hz). The numerical results of SPPs wave 
lengths also have a good precision. Fig.~\ref{fig:9} shows the slice of normalized SPPs field $H_{y}$ obtained by 
simulations, where the vacuum wave length $\lambda_{0}=240$ nm. The Z-projection of Fig.~\ref{fig:9} provide us with 
the SPPs mode structure on different sides of the silver-air interface. The localization of wave means subwavelength 
field structure and localized energy enhancement which can be used in many branches of nano-electrodynamics. Just as 
the first numerical experiment, the conserved quantities, such as the total Hamiltonian, in these simulations are still 
bounded by the discrete conservation laws.

The numerical experiments implemented in this work verify the numerical schemes and show the good properties of 
these schemes in secular simulations. It can be expected that this work will provide us with a powerful numerical tool 
for the first-principle based simulation study of plasmonics.

\section{Conclusion}
In this work, we have constructed a class of variational schemes for the hydrodynamic-electrodynamic model of lossless 
free electron gas in quasi-neutral background to implement high-quality simulations of the SPPs. The Lagrangian density 
of lossless free electron gas with a self-consistent electromagnetic field was established, and the conservation laws 
with constraints were obtained. By using the DEC-based discretization, we reconstructed a field theory on the discrete 
space-time manifold and constructed the variational schemes by discretizing and minimizing the action. The variational 
schemes are nonlinear semi-explicit, which means the nonlinear solvers are needed. We introduced a hybrid Newton-BICGSTAB 
method to solve the nonlinear algebraic equations involved in the variational schemes. Instead of discretizing the 
partial differential equations, the variational schemes have superior numerical properties in secular simulations, as 
they preserve the discrete Lagrangian symplectic structure, charge, and general energy-momentum density conservations. 
Two types of numerical experiments, i.e., bulk plasmon oscillation and 2-D SPPs, were implemented. The numerical results 
can reproduce characteristic dispersion relations of both types of plasmonic phenomena. The numerical errors of conserved 
quantities, e.g., the total Hamiltonian, in all examples are bounded by a small value after long-term simulations, which 
shows the advantages of variational schemes in secular simulations. The variational schemes constructed for the 
hydrodynamic-electrodynamic model can be used as a powerful numerical tool in plasmonics. Improvements, such as 
unstructured lattices, high-quality boundaries, more efficient algebraic solvers, and loss and interband transition of 
real metal, will be developed in our future work.

\begin{acknowledgments}
Q. Chen would like to thank H. Qin and J. Xiao for considerable help in differential geometry and field theory 
at University of Science and Technology of China. We also thank the anonymous reviewer, whose criticism and suggestions 
strongly improved this paper. This work is supported by the National Nature Science Foundation of China (NSFC-11805273, 
51477182), the CEMEE State Key Laboratory Foundation (CEMEE-2018Z0104B), and the Testing Technique Foundation for Young 
Scholars. Numerical simulations were implemented on the Shenma supercomputer at Institute of Plasma Physics, Chinese 
Academy of Sciences and the TH-1A supercomputer at National Super Computer Center in Tianjin.
\end{acknowledgments}

\appendix
\section{The Jacobian of Nonlinear Equations\label{sec:app}}
To implement the Newton-Raphson iteration, we should calculate the Jacobian of nonlinear equations in every 
iteration step. Here we give the detailed Jacobian elements. Assuming the nonlinear function $f_{1}(\bm{X}^{n+1})$ is 
defined by Eq.~\eqref{eq:38}, we obtain,
\begin{eqnarray}
\frac{\partial{f_{1}}}{\partial{\alpha^{n+\frac{3}{2}}_{i,j,k}}}=1,\label{eq:a1}
\end{eqnarray}
\begin{eqnarray}
\frac{\partial{f_{1}}}{\partial{v^{n+1}_{xi+\frac{1}{2},j,k}}}=-m\Delta{t}v^{n+1}_{xi+\frac{1}{2},j,k}-e\Delta{t}A^{n+1}_{xi+\frac{1}{2},j,k}+\frac{\Delta{t}}{\Delta{x}}\left(\alpha^{n+\frac{1}{2}}_{i+1,j,k}-\alpha^{n+\frac{1}{2}}_{i,j,k}\right),\label{eq:a2}
\end{eqnarray}
\begin{eqnarray}
\frac{\partial{f_{1}}}{\partial{v^{n+1}_{yi,j+\frac{1}{2},k}}}=-m\Delta{t}v^{n+1}_{yi,j+\frac{1}{2},k}-e\Delta{t}A^{n+1}_{yi,j+\frac{1}{2},k}+\frac{\Delta{t}}{\Delta{y}}\left(\alpha^{n+\frac{1}{2}}_{i,j+1,k}-\alpha^{n+\frac{1}{2}}_{i,j,k}\right),\label{eq:a3}
\end{eqnarray}
\begin{eqnarray}
\frac{\partial{f_{1}}}{\partial{v^{n+1}_{zi,j,k+\frac{1}{2}}}}=-m\Delta{t}v^{n+1}_{zi,j,k+\frac{1}{2}}-e\Delta{t}A^{n+1}_{zi,j,k+\frac{1}{2}}+\frac{\Delta{t}}{\Delta{z}}\left(\alpha^{n+\frac{1}{2}}_{i,j,k+1}-\alpha^{n+\frac{1}{2}}_{i,j,k}\right).\label{eq:a4}
\end{eqnarray}

Assuming the nonlinear function $f_{2}(\bm{X}^{n+1})$ is defined by Eq.~\eqref{eq:39}, we obtain,
\begin{eqnarray}
\frac{\partial{f_{2}}}{\partial{v^{n+1}_{xi+\frac{1}{2},j,k}}}=mn^{n+1}_{i+\frac{1}{2},j+\frac{1}{2},k+\frac{1}{2}},\label{eq:a5}
\end{eqnarray}
\begin{eqnarray}
\frac{\partial{f_{2}}}{\partial{n^{n+1}_{i+\frac{1}{2},j+\frac{1}{2},k+\frac{1}{2}}}}=mv^{n+1}_{xi+\frac{1}{2},j,k}+eA^{n+1}_{xi+\frac{1}{2},j,k}-\frac{\alpha^{n+\frac{1}{2}}_{i+1,j,k}-\alpha^{n+\frac{1}{2}}_{i,j,k}}{\Delta{x}},\label{eq:a6}
\end{eqnarray}
\begin{eqnarray}
\frac{\partial{f_{2}}}{\partial\lambda^{n+1}_{i+\frac{1}{2},j+\frac{1}{2},k+\frac{1}{2}}}=-\frac{\mu^{n+\frac{1}{2}}_{i+1,j,k}-\mu^{n+\frac{1}{2}}_{i,j,k}}{\Delta{x}}.\label{eq:a7}
\end{eqnarray}

Assuming the nonlinear function $f_{3}(\bm{X}^{n+1})$ is defined by Eq.~\eqref{eq:40}, we obtain,
\begin{eqnarray}
\frac{\partial{f_{3}}}{\partial{v^{n+1}_{yi,j+\frac{1}{2},k}}}=mn^{n+1}_{i+\frac{1}{2},j+\frac{1}{2},k+\frac{1}{2}},\label{eq:a8}
\end{eqnarray}
\begin{eqnarray}
\frac{\partial{f_{3}}}{\partial{n^{n+1}_{i+\frac{1}{2},j+\frac{1}{2},k+\frac{1}{2}}}}=mv^{n+1}_{yi,j+\frac{1}{2},k}+eA^{n+1}_{yi,j+\frac{1}{2},k}-\frac{\alpha^{n+\frac{1}{2}}_{i,j+1,k}-\alpha^{n+\frac{1}{2}}_{i,j,k}}{\Delta{y}},\label{eq:a9}
\end{eqnarray}
\begin{eqnarray}
\frac{\partial{f_{3}}}{\partial\lambda^{n+1}_{i+\frac{1}{2},j+\frac{1}{2},k+\frac{1}{2}}}=-\frac{\mu^{n+\frac{1}{2}}_{i,j+1,k}-\mu^{n+\frac{1}{2}}_{i,j,k}}{\Delta{y}}.\label{eq:a10}
\end{eqnarray}

Assuming the nonlinear function $f_{4}(\bm{X}^{n+1})$ is defined by Eq.~\eqref{eq:41}, we obtain,
\begin{eqnarray}
\frac{\partial{f_{4}}}{\partial{v^{n+1}_{zi,j,k+\frac{1}{2}}}}=mn^{n+1}_{i+\frac{1}{2},j+\frac{1}{2},k+\frac{1}{2}},\label{eq:a11}
\end{eqnarray}
\begin{eqnarray}
\frac{\partial{f_{4}}}{\partial{n^{n+1}_{i+\frac{1}{2},j+\frac{1}{2},k+\frac{1}{2}}}}=mv^{n+1}_{zi,j,k+\frac{1}{2}}+eA^{n+1}_{zi,j,k+\frac{1}{2}}-\frac{\alpha^{n+\frac{1}{2}}_{i,j,k+1}-\alpha^{n+\frac{1}{2}}_{i,j,k}}{\Delta{z}},\label{eq:a12}
\end{eqnarray}
\begin{eqnarray}
\frac{\partial{f_{4}}}{\partial\lambda^{n+1}_{i+\frac{1}{2},j+\frac{1}{2},k+\frac{1}{2}}}=-\frac{\mu^{n+\frac{1}{2}}_{i,j,k+1}-\mu^{n+\frac{1}{2}}_{i,j,k}}{\Delta{z}}.\label{eq:a13}
\end{eqnarray}

Assuming the nonlinear function $f_{5}(\bm{X}^{n+1})$ is defined by Eq.~\eqref{eq:46}, we obtain,
\begin{eqnarray}
\frac{\partial{f_{5}}}{\partial{v^{n+1}_{xi+\frac{1}{2},j,k}}}=\frac{\Delta{t}}{\Delta{x}}n^{n+1}_{i+\frac{1}{2},j+\frac{1}{2},k+\frac{1}{2}},\label{eq:a14}
\end{eqnarray}
\begin{eqnarray}
\frac{\partial{f_{5}}}{\partial{v^{n+1}_{xi-\frac{1}{2},j,k}}}=-\frac{\Delta{t}}{\Delta{x}}n^{n+1}_{i-\frac{1}{2},j+\frac{1}{2},k+\frac{1}{2}},\label{eq:a15}
\end{eqnarray}
\begin{eqnarray}
\frac{\partial{f_{5}}}{\partial{v^{n+1}_{yi,j+\frac{1}{2},k}}}=\frac{\Delta{t}}{\Delta{y}}n^{n+1}_{i+\frac{1}{2},j+\frac{1}{2},k+\frac{1}{2}},\label{eq:a16}
\end{eqnarray}
\begin{eqnarray}
\frac{\partial{f_{5}}}{\partial{v^{n+1}_{yi,j-\frac{1}{2},k}}}=-\frac{\Delta{t}}{\Delta{y}}n^{n+1}_{i+\frac{1}{2},j-\frac{1}{2},k+\frac{1}{2}},\label{eq:a17}
\end{eqnarray}
\begin{eqnarray}
\frac{\partial{f_{5}}}{\partial{v^{n+1}_{zi,j,k+\frac{1}{2}}}}=\frac{\Delta{t}}{\Delta{z}}n^{n+1}_{i+\frac{1}{2},j+\frac{1}{2},k+\frac{1}{2}},\label{eq:a18}
\end{eqnarray}
\begin{eqnarray}
\frac{\partial{f_{5}}}{\partial{v^{n+1}_{zi,j,k-\frac{1}{2}}}}=-\frac{\Delta{t}}{\Delta{z}}n^{n+1}_{i+\frac{1}{2},j+\frac{1}{2},k-\frac{1}{2}},\label{eq:a19}
\end{eqnarray}
\begin{eqnarray}
\frac{\partial{f_{5}}}{\partial{n^{n+1}_{i+\frac{1}{2},j+\frac{1}{2},k+\frac{1}{2}}}}=1+\frac{\Delta{t}}{\Delta{x}}v^{n+1}_{xi+\frac{1}{2},j,k}+\frac{\Delta{t}}{\Delta{y}}v^{n+1}_{yi,j+\frac{1}{2},k}+\frac{\Delta{t}}{\Delta{z}}v^{n+1}_{zi,j,k+\frac{1}{2}},\label{eq:a20}
\end{eqnarray}
\begin{eqnarray}
\frac{\partial{f_{5}}}{\partial{n^{n+1}_{i-\frac{1}{2},j+\frac{1}{2},k+\frac{1}{2}}}}=-\frac{\Delta{t}}{\Delta{x}}v^{n+1}_{xi-\frac{1}{2},j,k},\label{eq:a21}
\end{eqnarray}
\begin{eqnarray}
\frac{\partial{f_{5}}}{\partial{n^{n+1}_{i+\frac{1}{2},j-\frac{1}{2},k+\frac{1}{2}}}}=-\frac{\Delta{t}}{\Delta{y}}v^{n+1}_{yi,j-\frac{1}{2},k},\label{eq:a22}
\end{eqnarray}
\begin{eqnarray}
\frac{\partial{f_{5}}}{\partial{n^{n+1}_{i+\frac{1}{2},j+\frac{1}{2},k-\frac{1}{2}}}}=-\frac{\Delta{t}}{\Delta{z}}v^{n+1}_{zi,j,k-\frac{1}{2}}.\label{eq:a23}
\end{eqnarray}

Assuming the nonlinear function $f_{6}(\bm{X}^{n+1})$ is defined by Eq.~\eqref{eq:48}, we obtain,
\begin{eqnarray}
\frac{\partial{f_{6}}}{\partial{v^{n+1}_{xi+\frac{1}{2},j,k}}}=\frac{\Delta{t}}{\Delta{x}}\lambda^{n+1}_{i+\frac{1}{2},j+\frac{1}{2},k+\frac{1}{2}},\label{eq:a24}
\end{eqnarray}
\begin{eqnarray}
\frac{\partial{f_{6}}}{\partial{v^{n+1}_{xi-\frac{1}{2},j,k}}}=-\frac{\Delta{t}}{\Delta{x}}\lambda^{n+1}_{i-\frac{1}{2},j+\frac{1}{2},k+\frac{1}{2}},\label{eq:a25}
\end{eqnarray}
\begin{eqnarray}
\frac{\partial{f_{6}}}{\partial{v^{n+1}_{yi,j+\frac{1}{2},k}}}=\frac{\Delta{t}}{\Delta{y}}\lambda^{n+1}_{i+\frac{1}{2},j+\frac{1}{2},k+\frac{1}{2}},\label{eq:a26}
\end{eqnarray}
\begin{eqnarray}
\frac{\partial{f_{6}}}{\partial{v^{n+1}_{yi,j-\frac{1}{2},k}}}=-\frac{\Delta{t}}{\Delta{y}}\lambda^{n+1}_{i+\frac{1}{2},j-\frac{1}{2},k+\frac{1}{2}},\label{eq:a27}
\end{eqnarray}
\begin{eqnarray}
\frac{\partial{f_{6}}}{\partial{v^{n+1}_{zi,j,k+\frac{1}{2}}}}=\frac{\Delta{t}}{\Delta{z}}\lambda^{n+1}_{i+\frac{1}{2},j+\frac{1}{2},k+\frac{1}{2}},\label{eq:a28}
\end{eqnarray}
\begin{eqnarray}
\frac{\partial{f_{6}}}{\partial{v^{n+1}_{zi,j,k-\frac{1}{2}}}}=-\frac{\Delta{t}}{\Delta{z}}\lambda^{n+1}_{i+\frac{1}{2},j+\frac{1}{2},k-\frac{1}{2}},\label{eq:a29}
\end{eqnarray}
\begin{eqnarray}
\frac{\partial{f_{6}}}{\partial\lambda^{n+1}_{i+\frac{1}{2},j+\frac{1}{2},k+\frac{1}{2}}}=1+\frac{\Delta{t}}{\Delta{x}}v^{n+1}_{xi+\frac{1}{2},j,k}+\frac{\Delta{t}}{\Delta{y}}v^{n+1}_{yi,j+\frac{1}{2},k}+\frac{\Delta{t}}{\Delta{z}}v^{n+1}_{zi,j,k+\frac{1}{2}},\label{eq:a30}
\end{eqnarray}
\begin{eqnarray}
\frac{\partial{f_{6}}}{\partial\lambda^{n+1}_{i-\frac{1}{2},j+\frac{1}{2},k+\frac{1}{2}}}=-\frac{\Delta{t}}{\Delta{x}}v^{n+1}_{xi-\frac{1}{2},j,k},\label{eq:a31}
\end{eqnarray}
\begin{eqnarray}
\frac{\partial{f_{6}}}{\partial\lambda^{n+1}_{i+\frac{1}{2},j-\frac{1}{2},k+\frac{1}{2}}}=-\frac{\Delta{t}}{\Delta{y}}v^{n+1}_{yi,j-\frac{1}{2},k},\label{eq:a32}
\end{eqnarray}
\begin{eqnarray}
\frac{\partial{f_{6}}}{\partial\lambda^{n+1}_{i+\frac{1}{2},j+\frac{1}{2},k-\frac{1}{2}}}=-\frac{\Delta{t}}{\Delta{z}}v^{n+1}_{zi,j,k-\frac{1}{2}}.\label{eq:a33}
\end{eqnarray}
It can be seen that the Jacobian $\bm{J}_{F}$ is a large sparse matrix in every iteration step.

\nocite{*}

\end{document}